\documentclass[fleqn,usenatbib]{mnras}

\usepackage{newtxtext,newtxmath}

\newcommand{\mh}{\textcolor{red}}

\usepackage[T1]{fontenc}

\DeclareRobustCommand{\VAN}[3]{#2}
\let\VANthebibliography\thebibliography
\def\thebibliography{\DeclareRobustCommand{\VAN}[3]{##3}\VANthebibliography}

\usepackage{graphicx}	
\usepackage{amsmath}	
\usepackage{subcaption}
\usepackage[normalem]{ulem}

\newcommand{\numax}{\nu_{\text{max}}}

\newcommand{\Teff}{T_{\text{eff}}}
\newcommand{\Teffsun}{\mathrm{T}_{\text{eff},\odot}}

\newcommand{\msun}{\mathrm{M}\textsubscript{\(\odot\)}}

\newcommand{\Lsun}{\mathrm{L}\textsubscript{\(\odot\)}}

\defcitealias{Howell22_M4}{HowellM4}
\defcitealias{Howell_M80}{HowellM80}

\title[Asteroseismology of Red Giants in M9 \& M19]{Asteroseismic Masses of Red Giants in the Galactic Globular Clusters M9 \& M19}

\author[M. Howell et al.]{
Madeline Howell,$^{1,2}$\thanks{E-mail: madeline.howell1@monash.edu}
Simon W. Campbell,$^{1,2}$
Csilla Kalup,$^{3,4}$
Dennis Stello$^{5,6,2}$
and Gayandhi M. De Silva$^{7,2}$
\\
$^{1}$School of Physics and Astronomy, Monash University, Clayton, VIC 3800, Australia\\
$^{2}$ARC Centre of Excellence for Astrophysics in Three Dimensions (ASTRO-3D), Australia\\
$^{3}$Konkoly Observatory, HUN-REN Research Centre for Astronomy and Earth Sciences, MTA Centre of Excellence, Konkoly-Thege Mikl\'os \'ut 15-17, H-1121, \\ 
Budapest, Hungary\\
$^{4}$ELTE E\"otv\"os Lor\'and University, Institute of Physics and Astronomy, 1117, P\'azm\'any P\'eter s\'et\'any 1/A, Budapest, Hungary\\
$^{5}$School of Physics, University of New South Wales, NSW 2052, Sydney, Australia\\
$^{6}$Sydney Institute for Astronomy (SIfA), School of Physics, University of Sydney, NSW 2006, Sydney, Australia\\
$^{7}$School of Mathematical and Physical Sciences, Faculty of Science and Engineering, Macquarie University, Macquarie Park, NSW 2113, Australia\\
}

\date{Accepted XXX. Received YYY; in original form ZZZ}

\pubyear{2024}

\begin{document}
\label{firstpage}
\pagerange{\pageref{firstpage}--\pageref{lastpage}}
\maketitle

\begin{abstract}
Asteroseismic masses of globular cluster (GC) stars are invaluable to investigate stellar evolution. Previously, only two GCs have been seismically studied. We present new detections of solar-like oscillations in the clusters M9 and M19, focusing on two key areas: stellar mass loss and GC multiple populations. Using K2 photometry, we detect solar-like oscillations in stars on the red giant branch and early asymptotic giant branch. We measure an integrated mass-loss for M9 of $0.16\pm0.02$(rand){\raisebox{0.5ex}{\tiny$\substack{+0.03 \\ -0.03}$}}(sys)$~\msun$ and M19 of $0.33\pm0.03$(rand){\raisebox{0.5ex}{\tiny$\substack{+0.09 \\ -0.07}$}}(sys)$~\msun$. Comparing these to the mass-loss estimates from previous seismically studied clusters, we derive a preliminary relationship between stellar mass-loss and metallicity for Type I GCs. We find that the mass-loss for M19 -- a Type II GC-- is significantly larger, suggesting Type II clusters follow a different mass-loss-metallicity trend.  We also examine the mass distributions in each evolutionary phase for evidence of a bimodality that could indicate mass differences between sub-populations. While no clear bimodality is observed, there is tentative evidence suggesting the presence of two mass populations. Classification through spectroscopic abundances into the sub-populations is needed to verify these findings. This study reinforces that asteroseismology of GC stars provides an excellent testbed for studying stellar evolution. However, to advance the field we need high-quality photometry of more GCs, a goal that could be realised with the upcoming \textit{Roman Telescope}. 
\end{abstract}




\section{Introduction}
\label{sec:intro}


Globular clusters (GCs) are excellent laboratories for benchmarking stellar evolution due to their relatively homogeneous samples of stars. Quantifying stellar masses is essential because mass is a key defining factor in the evolution of a star. Asteroseismology can be used to accurately estimate stellar masses through the measurement of solar-like oscillations. Measuring asteroseismic masses of GC stars is therefore one of the most advantageous tools we have to investigate low mass stellar evolution.

The first asteroseismic studies of GCs used the \textit{Hubble Space Telescope} \citep{Edmonds96_Tuc47asteroseismology, Stello09_HST_GC_seismo} and ground-based photometry \citep{Frandsen07_M4asteroseismology} to attempt to measure the pressure modes of stars in the GCs 47 Tuc, NGC 6397, and M4. In all these studies, only marginal detections of solar-like oscillations were found and no useful measurements of the global seismic parameters ($\Delta\nu$ \& $\numax$) were made. They concluded that to be able to detect solar-like oscillations in GC stars, higher frequency resolution and signal-to-noise was needed.

In more recent efforts, asteroseismology has been successfully used to measure accurate masses of red giant branch (RGB), horizontal branch (HB) and asymptotic giant branch (AGB) stars in two GCs using photometry from the \textit{K2} mission \citep{Howell14_K2_mission}. \citet{Miglio_M4_study} was the first study with concrete detections of solar-like oscillations in the GC M4, with the measurement of seismic masses in 8 red giants. Subsequently, two more studies have increased the number of seismically studied red giants in M4 to 37 in \citet{Tailo22_M4} and 75 in \citet[][hereby \citetalias{Howell22_M4}]{Howell22_M4}. Solar-like oscillations of red giants have also been detected in the fainter and more metal-poor GC M80 \citet[][hereby \citetalias{Howell_M80}]{Howell_M80}. In this paper, we expand upon these studies by measuring seismic masses of red giants in two more GCs with \textit{K2} photometry: M9 (NGC 6333) and M19 (NGC 6273). 

M9 can be described as a `typical' GC with homogeneous heavy-element compositions ([Fe/H]~$=-1.67\pm0.01(\text{statistical})\pm0.19(\text{sys})$; \citealt{Arellano13_M9FeH}), and as such is categorised as a Type I cluster. In contrast, M19 belongs to a different class of GCs known as Type II or `iron-complex' clusters \citep[see e.g.][]{DaCosta16_ironcomplex,Marino15_ironcomplex,McKenzie22_ironcomplex}. These types of clusters exhibit large variations in iron and neutron-capture element abundances, where M19 was measured to have a maximum spread in [Fe/H] of $\sim1$~dex \citep{Johnson15_M19, Yong_2016_M19,Johnson_2017_M19}. Type II clusters are strongly suspected to be the stripped cores of former dwarf galaxies \citep[e.g.][]{Olszewski09_strippeddwarfgalaxy,Kuzma16_strippeddwarfgalaxy,DaCosta16_ironcomplex} accreted into the Milky Way. Thus, they could be used to trace the Galaxy's formation history.  

Previous photometric studies of M9 and M19 have measured the pulsations of variable stars such as RR Lyrae stars and Cepheids (e.g. \citealt{Sawyer1948,Clement1984,Clement01_VariablesM19,Clement1999, Arellano13_M9FeH,Ngeow22_ZwickyVariables}). However, there are no attempts at measuring solar-like oscillations in evolved stars in the literature. The objectives for our asteroseismic study of M9 and M19 are the same as \citetalias{Howell22_M4} and \citetalias{Howell_M80}; we aim to increase the number of GC stars with detected solar-like oscillations, estimate the individual stellar masses, measure an integrated stellar mass loss, and attempt to detect a mass difference between the sub-populations in the GCs. We summarise our science aims here and refer the reader to \citetalias{Howell22_M4} and \citetalias{Howell_M80} for an in-depth introduction. 

For low mass stars, mass loss is most significant during the RGB evolutionary phase. The amount of matter lost will impact the subsequent evolution of the star, i.e. on the HB and AGB. An integrated mass loss can be quantified by taking the differences in the averaged masses of various evolutionary phases (e.g. \citealt{Miglio12_OCstudy}; \citealt{Tailo22_M4}; \citetalias{Howell22_M4}; \citetalias{Howell_M80}). As detailed in \citetalias{Howell_M80}, stellar mass loss is thought to be metallicity dependent. By measuring the integrated mass loss for a sample of GCs, we can test whether higher-metallicity stars lose more mass compared to lower-metallicity stars. There is already promising evidence of this mass loss-metallicity trend from the comparison between the M4 ([Fe/H]~$ =-1.1$; \citealt{M4_metallicity2}) and M80 ([Fe/H]~$ =-1.8$; \citealt{Caretta15_M80}) seismic mass studies. Two additional seismically measured mass loss estimates at different metallicities -- as provided in this paper-- can help refine this relationship. 

We can also use the distribution of asteroseismic masses within each cluster to test if there is a mass difference between GC sub-populations. There is observational evidence that GCs contain typically two sub-populations -- sometimes more -- which vary in light elemental abundances (e.g. C, N, O, Na, O, and He; \citealt{Sneden1999_light_elemental_abundances1, Gratton12_light_elemental_abundances2}). Mass differences between the sub-populations are thought to occur due to the variation in He abundances, as shown in \citet{MacLean18_chloes_paper} and \citet{Jang19_2pops_models}. \citetalias{Howell_M80} was the first study to report a tentative detection of a mass difference between sub-populations in the GC M80, where there is clear evidence of a bimodal mass distribution in their early AGB sample. Furthermore, the peaks of this bimodal distribution are consistent with the sub-population masses after RGB mass loss in the independent modelling study by \citet{Tailo20_massloss_difference_multipops}. However, this result needs to be confirmed with spectroscopic chemical abundances to classify their sample into sub-populations. If confirmed, this could have implications for our understanding of the formation of GCs, and also long-standing mysteries of GCs such as the second parameter problem \citep[see for further details][]{VanDenBerg1967_2ndparamproblem,Sandage1967_2ndparam,FusiPecci1993_2ndparamproblem}.

This paper is set out as follows: Section 2 details our method of data preparation and analysis. In Sections 3 \& 4 we explain how we measured the global asteroseismic parameters and stellar parameters, respectively. We present our mass results and discuss their implications in Section 5, and provide a conclusion in Section 6.

\section{Data Preparation and Analysis}
\label{sec:data_prep}
\subsection{Cluster Superstamps}
\label{sec:superstamps}
\begin{figure*}
	\centering
	\includegraphics[width=2.1\columnwidth]{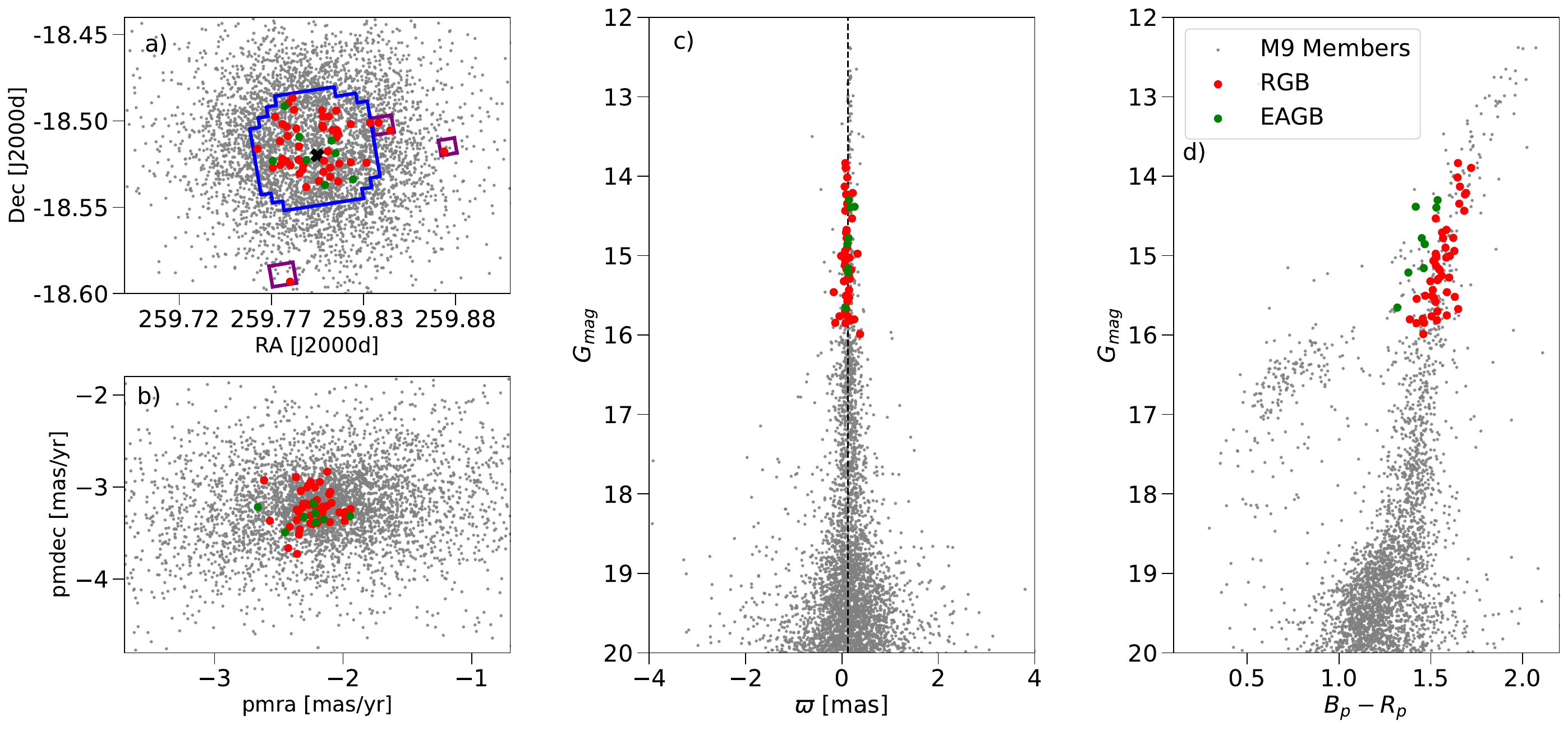}
	\caption{\textbf{(a) } The spatial positions of our M9 sample (coloured points) and a full \textit{Gaia} EDR3 membership sample \citep[grey;][]{Vasiliev21_GalacticGC_memberships}. Stars for which we were able to detect solar-like oscillations are classified into two evolutionary phases; RGB and EAGB, and are indicated by the larger coloured points. We also show the M9 \textit{K2} superstamp (blue) and the stars with their own TPFs (purple). The black cross indicates the centre of the cluster. \textbf{(b) } \textit{Gaia} DR3 proper motions of the same sample. \textbf{(c) } The \textit{Gaia} DR3 parallaxes, $\varpi$, for the cluster members and our seismic sample. The average parallax ($\varpi = 0.12$) for the cluster is indicated by the dashed line. \textbf{(d)} Gaia DR3 colour-magnitude diagram of our sample}
	\label{fig:CMD_M9}
\end{figure*}

\begin{figure*}
	\centering
	\includegraphics[width=2.1\columnwidth]{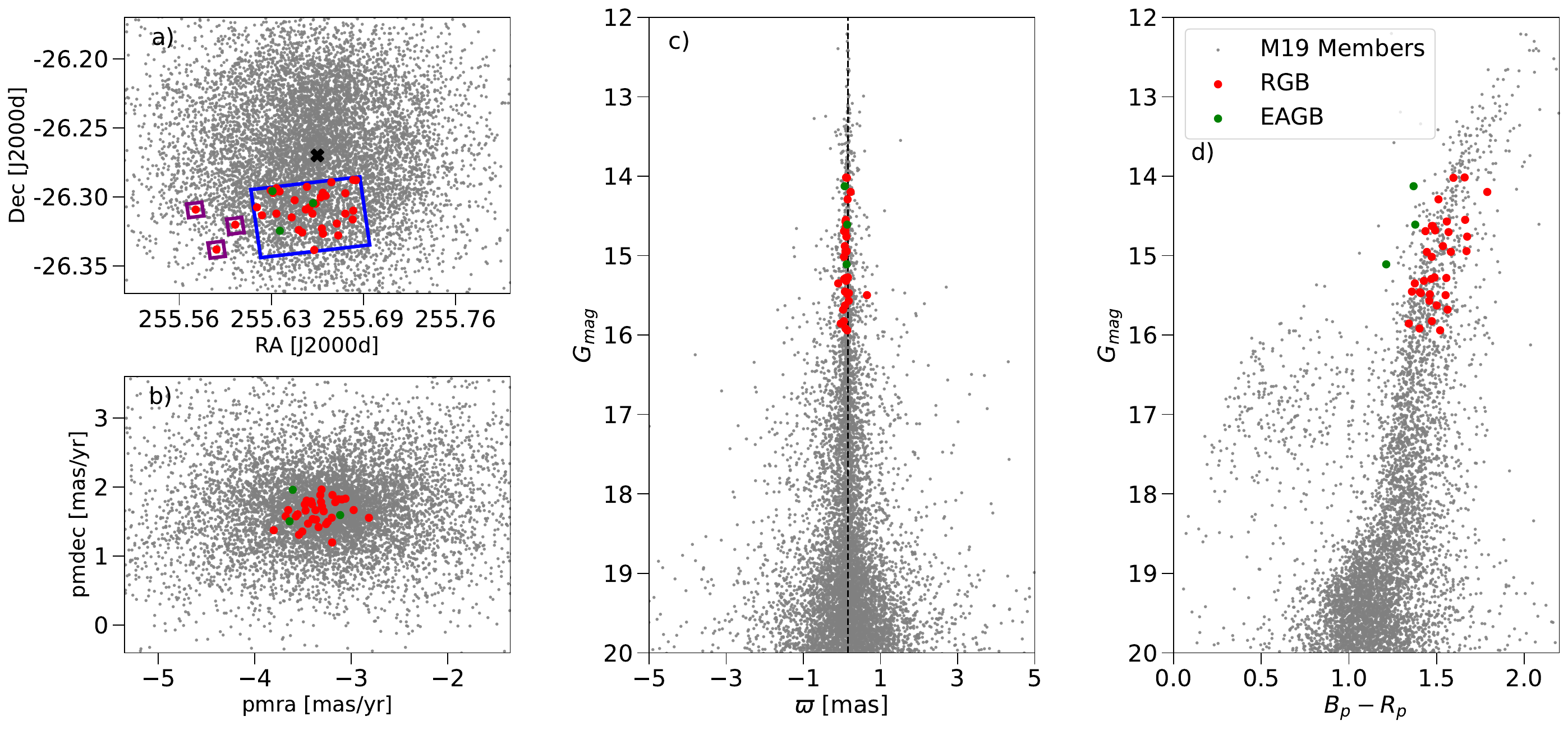}
	\caption{Same as Fig.~\ref{fig:CMD_M9} but for the cluster M19. The average parallax for this cluster is $\varpi = 0.15$}
	\label{fig:CMD_M19}
\end{figure*}

We used photometry from Campaign 11 of the \textit{K2} mission \citep{Howell14_K2_mission} for both clusters. The \textit{K2} superstamp of M9 captured the cluster centre (Fig.~\ref{fig:CMD_M9}a), whereas the superstamp of M19 observed a small rectangular field of view of the outskirts of the cluster (Fig.~\ref{fig:CMD_M19}a). Each cluster superstamp consisted of 127 (M9) and 116 (M19) individual target pixel files (TPFs) in grids of 5x5 pixels. 

\begin{figure}
	\centering
	\includegraphics[width=1\columnwidth]{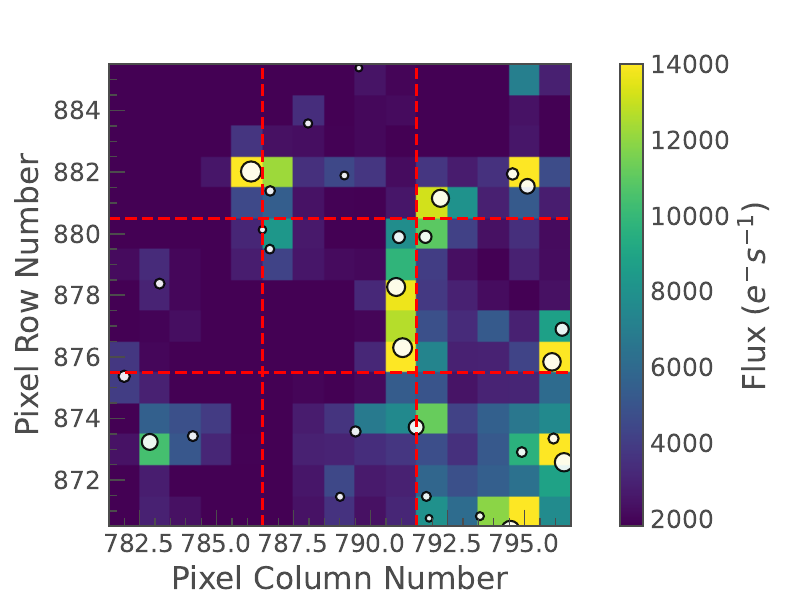}
	\caption{An example TPF patch for M9 from \texttt{TPFstitch}. The red lines divide the patch into the 9 individual TPFs (EPIC 200145500-200145502, 200145512-200145514, \& 200145488-200145490). The centre of the stars contained in the field are illustrated by the white circles, where the larger circles indicate stars with brighter magnitudes.}
	\label{fig:tpfstitch_eg}
\end{figure}

Due to the small 5x5 pixel size of the TPF data products, there was a significant chance that a potential target could be located on the edge of more than one TPF. We are unable to define an aperture mask (see Sec.~\ref{sec:Ap_masks} for the description of the method) for a single star when it is contained in several TPFs. Hence, we developed a Python function to `stitch' neighbouring TPFs together, which we call \texttt{TPFstitch}\footnote{The TPFstitch repository can be accessed at: \href{https://github.com/maddyhowell/TPFstitch}{https://github.com/maddyhowell/TPFstitch}}. By inputting the \texttt{FITs} files of between 2-9 TPFs that share a border, \texttt{TPFstitch} produces a single \texttt{FITs} file containing the stitched TPFs, which we refer to as a TPF `patch' (see example in Fig.~\ref{fig:tpfstitch_eg}). The world coordinate system is maintained within the final product, i.e. the RAs and DECs of targets can be identified in the resultant TPF patch. Additionally, the patch \texttt{FITs} file can be read straight into asteroseismic data analysis tools, such as \texttt{lightkurve} \citep{Lightkurve}. Currently, \texttt{TPFstitch} is only designed for the M9 and M19 clusters, but it could be easily adapted for any \textit{K2} object with several TPFs; e.g. other clusters, asteroids/comets, microlensing events, galaxies, etc. There is also the potential for \texttt{TPFstitch} to be applied to other photometric surveys, such as \textit{TESS} \citep{Ricker15_TESS}. 

We show one frame of the final cluster Superstamps (i.e. the full collective of patches) in Appendix~\ref{sec:Appendix_A} for M9 and Appendix~\ref{sec:Appendix_B} for M19. 

\subsection{Target Sample Selection}
\label{sec:target_selection}
Our target sample was selected using the Gaia DR3 source catalog \citep{Gaia_mission_paper, Gaia_DR3}. To ensure that an asteroseismic signal could be detected above the \textit{K2} noise level, we set a cut-off magnitude in the \textit{Gaia} G-band of $16$~mag (based on the results from \citetalias{Howell_M80}). A cluster membership probability for each star was obtained from \citet{Vasiliev21_GalacticGC_memberships}, using proper motions and parallaxes to determine membership (see Fig~\ref{fig:CMD_M9}b,c for M9 and Fig~\ref{fig:CMD_M19}b,c for M19). We accepted a membership probability greater than 0.8, however 83\% of our final membership sample had probabilities of $>0.99$. Next, we identified which stars have photometric data from the \textit{K2} mission using \texttt{K2fov} \citep{K2FOV}. We also searched around each cluster superstamp for individual TPFs that could contain cluster members. We provide the list of EPIC IDs associated with the TPFs used in this study in Table~\ref{tab:EPIC_IDs}. Star IDs were assigned following a method similar to \citetalias{Howell22_M4}.


\begin{table}
\centering
\footnotesize
\caption{The \textit{K2} EPIC IDs for the TPFs used in this study for M9 \& M19. Campaign 11 was split into two parts, C11a and C11b (see Sec.~\ref{sec:Ap_masks} for more details). Thus, there are separate Superstamps for each sub-campaign. Some stars were contained in separate TPFs to the cluster Superstamps, as indicated by individual EPIC IDs. }
    \begin{tabular}{cc}
      \hline 
      Object & EPIC ID(s) \\
      \hline 
      \textbf{M9} & \\
      C11a Superstamp & 200126982 - 200127109\\
      C11b Superstamp  & 200145408 - 200145535\\
      M9RGB164  & 234559535 \\
      M9RGB292 \& M9RGB41 & 251248689 \\
      M9RGB295 & 234572782 \\
      \hline
      \textbf{M19} & \\
      C11a Superstamp & 200126870 - 	200126981  \\
      C11b Superstamp  & 200145291 - 200145407  \\
      M19RGB261  & 232311226 \\
      M19RGB352 &  232307616\\
      M19RGB355 & 232313427 \\
      \hline 
    \end{tabular}
\label{tab:EPIC_IDs}
\end{table}

The evolutionary status for each star was determined using UBVRI photometry from \citet{Stetson19_GCS_UBVRI_photometry}. It has been shown (e.g. \citealt{MacLean18_chloes_paper}) that by using the colour combinations of V-(B-V) and U-(U-I), the AGB clearly separates from the RGB, and as such makes it easy to distinguish between these evolutionary phases. Unfortunately, the limiting $G$-magnitude of 16 means that HB stars were too faint to be included in the target sample for both clusters. Therefore, our red giant sample only consists of early AGB (EAGB; the phase of evolution directly after the HB) and RGB stars.



\subsection{Photometric Light Curves \& Power Spectra}
\label{sec:Ap_masks}
Campaign 11 was split into two parts, labelled as C11a and C11b. This was due to an error in the initial roll angle used to minimize solar torque on the spacecraft. Campaign 11 had a total observation length of $\sim 69$ days, split into $\sim 22$ days for C11a and $\sim 42$ days for C11b. There was a five day gap between the sub-campaigns, during which the roll angle was corrected. The photometry from each sub-campaign was initially analysed separately. 

The time series flux data was extracted using aperture masks, where we implemented the same method as \citetalias{Howell22_M4} and \citetalias{Howell_M80}. To summarise, the aperture masks were found by choosing the brightest pixels within a certain threshold of the identified central pixel of the star. The same aperture mask for a star was used for both sub-campaigns. Due to the high crowding of GCs, we took care to not include any pixels in the aperture masks that contained neighbouring stars that could potentially contaminate the photometry. To detrend the photometric data, we again used the same method as \citetalias{Howell22_M4} and \citetalias{Howell_M80}. This involved a two-step process: first applying a pixel-level decorrelator to the target pixel file (based on the method by \citealt{Deming15_PLD_creator}) to remove systematic effects from individual pixels (e.g. instrumental drifts), and then removing long-term trends from the photometric light curves by using a self-flat fielding corrector \citep{Vanderburg14_SFF_cleaning}. We refer to Section 2.2 of \citetalias{Howell22_M4} for a detailed description of the methodology. In this study we also filled short gaps in the data if there were 3 or less missing consecutive cadences, following the method in \citet{Stello2015}.

Once we had obtained cleaned light curves, we made a Lomb-Scargle power spectral density periodogram \citep{Lomb,Scargle} for each star. We searched these spectra for a power excess which could indicate the presence of solar-like oscillations, using a prediction for $\numax$ to guide the eye. To predict an estimate for $\numax$, we derived a relationship between Gaia $G$-magnitude and the measured $\numax$ from a sample of M80\footnote{M80 has a similar distance modulus as both M9 and M19, and thus this derived relationship is a good initial estimate for each star.} stars \citepalias{Howell_M80}: $\numax = 1.229\times 10^{-19}G_{\rm mag}^{16.89}$. If a power excess near the predicted $\numax$ could not be seen, the star was removed from the target sample.

Due to the shorter observation length of C11a and the wrong roll angle, we tested whether the photometric quality was similar to C11b by comparing the white noise metrics (calculated as the median power density between $260-280~\mu\mathrm{Hz}$; \citealt{Stello2015}). We also checked whether combining the light curves from each sub-campaign reduced the white noise metric. The large 5-day gap between the sub-campaigns was not filled. We found that for the majority of the sample, the white noise metric was the smallest when combining the light curves from both sub-campaigns. However for a small sample of stars, no signal was seen when  combining the light curves, but was observed in the individual sub-campaign spectra. For these cases, we chose the light curve from the campaign with the lowest white noise metric. In Table~\ref{tab:final_results_M9} and \ref{tab:final_results_M19}, we have indicated whether the full campaign 11 is used or if only one sub-campaign was used for each star. 

Due to the large pixel sizes of \textit{K2}'s CCDs (4"/pixel), we wanted to use an objective method to confirm that there was no photometric contamination in our light curves. We performed a final contamination check by calculating a similarity metric with neighbouring stars. We adopted the `confusion from blends' metric introduced in \citet{Stello22_tess_red_giants}, which uses the Shape-Base Distance (SBD) algorithm \citep{Paparrizos_2016_kshapes} to quantify the similarity between two power spectra. We implemented this similarity metric whenever there was a nearby star within 1 magnitude or brighter than the target star and within 4 pixels of the centroid pixel using the following method:

\begin{enumerate}
    \item determine an estimate of $\numax$ for the neighbouring star, by finding the apparent $G$-magnitude of the star at the distance of the cluster. The initial distance of the neighbouring star was determined from the Gaia parallax.
    \item if the $\numax$ estimate for the neighbouring star is within 10$\mu\mathrm{Hz}$ of the predicted $\numax$ for the target star, we follow the method outlined above to create a light curve and the Lomb-Scargle power spectral density periodogram for the neighbouring star. If the $\numax$ estimate for the neighbouring star is not similar to the predicted target star's $\numax$, we assume that it is highly unlikely there is photometric contamination from this star, and we do not go any further.
    \item For potential contaminators, we calculated the `confusion from blends' metric between the target and neighbour. 
    \item If the similarity metric was less than 0.5 (0 corresponds to spectra being the same and 2 corresponds to dissimilar spectra), we concluded that there could be photometric contamination occurring and the target star was removed from the sample. 
\end{enumerate}

\section{Global Asteroseismic Quantities}
\label{sec:seismic_params}
Similar to \citetalias{Howell22_M4} and \citetalias{Howell_M80}, we concluded that the signal-to-noise and frequency resolution of our power spectra is too low to estimate an accurate $\Delta\nu$, and therefore for each star we only measure $\numax$. As demonstrated by previous asteroseismic GC studies, estimates of $\numax$, temperature, and luminosity for each star is sufficient to measure accurate asteroseismic masses. \citetalias{Howell22_M4} and \citetalias{Howell_M80} both used \texttt{pySYD} \citep{pySYD_chontos} for the global seismic quantity estimation. However, this pipeline cannot solely measure $\numax$ without also measuring $\Delta\nu$, and is dependent on the measured $\Delta\nu$ to smooth the power spectrum to determine $\numax$. 

\begin{figure}
	\centering
	\includegraphics[width=\columnwidth]{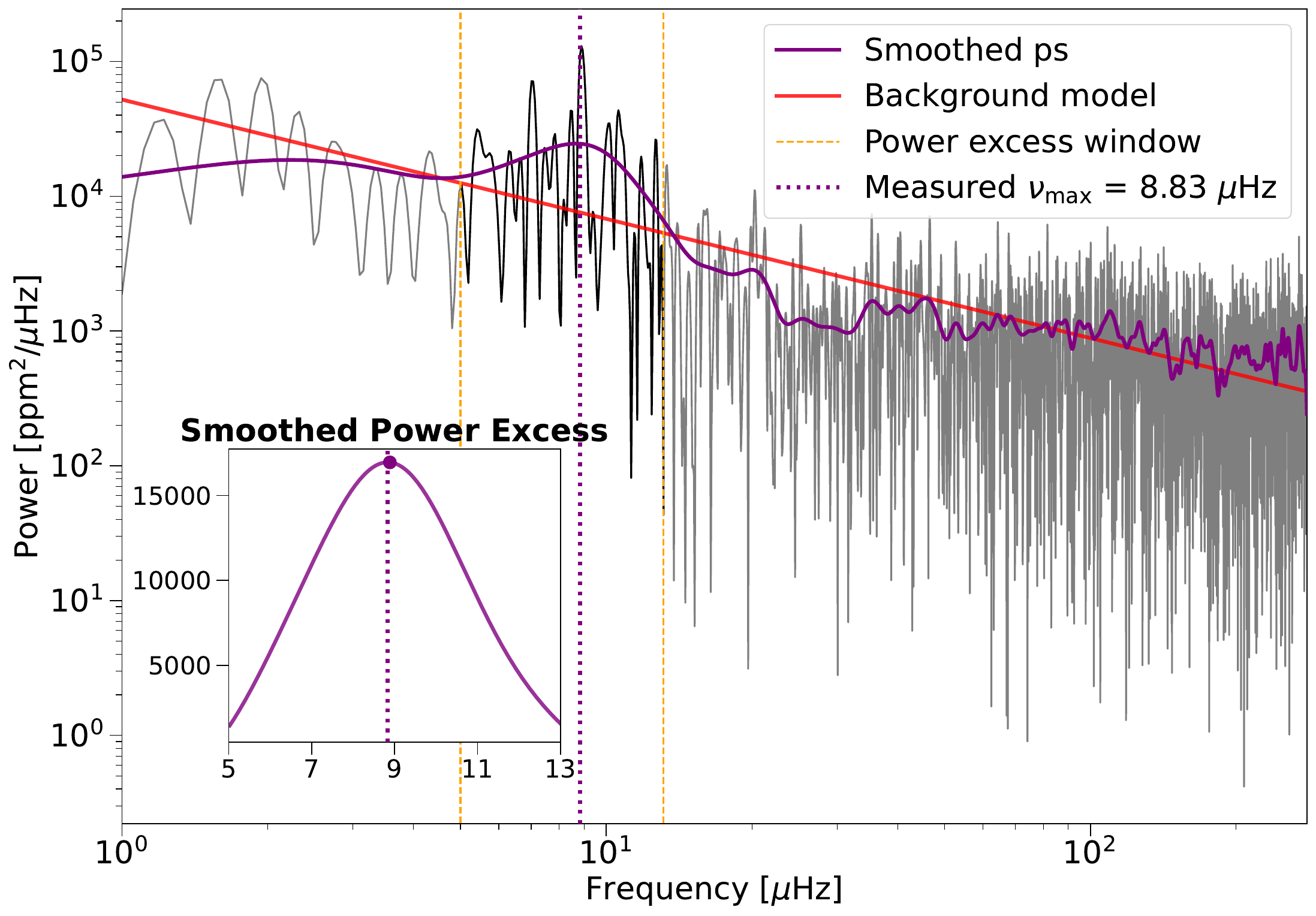}
	\caption{Power spectrum for M9RGB289 (grey), smoothed power spectrum (purple), and the linear background fit (red). The power excess of the solar-like oscillation (black) is situated between the two orange lines. \texttt{pyMON} estimates $\numax$ as the frequency of the maximum power in the background corrected smoothed power spectrum (as shown in the inset).}
	\label{fig:ps_eg}
\end{figure}

We developed our own code \texttt{pyMON}\footnote{You can access the pyMON repository here: \href{https://github.com/maddyhowell/pyMON}{https://github.com/maddyhowell/pyMON}}, directly based on \texttt{pySYD}, to only estimate $\numax$ and its associated uncertainty. Similar to \citetalias{Howell22_M4} and \citetalias{Howell_M80}, we remove the background from the power spectrum by subtracting a linear model estimated from the intersection between the power excess boundaries (see Fig.\ref{fig:ps_eg} for an example). After background removal, the power spectrum is heavily smoothed based on a $\Delta\nu$ estimate from a $\Delta\nu$-$\numax$ scaling relation (see Sec.3 of \citetalias{Howell22_M4} for specific relations, and \citealt{Stello09_dnu_numax_relation} for more details), and the frequency for the maximum amplitude in the power excess window is adopted as the measured $\numax$. To find the uncertainty, the power spectrum is perturbed with stochastic noise and a new estimate for $\numax$ is found (refer to \citealt{Huber09_SYDpipeline} for a  description of this method). This is repeated 500 times. The standard deviation of the resulting $\numax$ measurement distribution is adopted as the uncertainty. 

For a sub-sample of RGB stars from both M9 and M19, we calculated $\numax$ from both pipelines. We found that the $\numax$ estimates were consistent between the pipelines, although the \texttt{pyMON} uncertainties were approximately a factor of two smaller than the calculated uncertainties from \texttt{pySYD}. The dependency of the measured $\Delta\nu$ to smooth the power spectrum directly affects the measurement of $\numax$. Hence, we speculate that bad measurements of $\Delta\nu$ will inflate the $\numax$ uncertainty in \texttt{pySYD}. As such, our method of using a scaling relation to estimate $\Delta\nu$ results in a $\numax$ uncertainty that is more robust for data that has low signal-to-noise and frequency resolution. This was also evidenced by a comparison to the large asteroseismic survey K2GAP \citep{Zinn20_K2GAP_DR2,Zinn21_K2GAP_DR3}. Our average  $\numax$ fractional uncertainty using \texttt{pyMON} was $\sim2.5\%$, which is slightly larger than the $\numax$ fractional uncertainty of $\sim1\%$ for RGB stars in the K2GAP survey. In contrast, \citetalias{Howell22_M4} and \citetalias{Howell_M80} reported average $\numax$ fractional uncertainties between $4.5-6.33\%$ when using pySYD.

In total, we detected solar-like oscillations in 55 red giants in M9 (coloured points in Fig.~\ref{fig:CMD_M9}) and 37 red giants in M19 (coloured points in Fig.~\ref{fig:CMD_M19}). Each star is assigned a quality flag: marginal detection (MD) or detection (D) following the same criteria in \citetalias{Howell22_M4} and \citetalias{Howell_M80}. The estimates for $\numax$ and the quality flag for each star in our sample is given in Table~\ref{tab:final_results_M9} for M9 and Table~\ref{tab:final_results_M19} for M19. 

\section{Stellar and Cluster Properties}
\label{sec:stellar_parameters}
\subsection{Dust corrections, isochrones \& distance modulus}
\label{sec:dist_mod}
The distances to GCs are typically measured using three techniques; isochrone fits to colour-magnitude diagrams \citep[e.g.][]{Ferraro1999, Dotter11,Valcin20}, parallax measurements (e.g. from Gaia \citep{Gaia_mission_paper}) or from the periods and absolute luminosities of variable stars such as RR Lyraes, Cepheids and Mira-types \citep[e.g.][]{Bono07, Matsunaga06, Feast02}. \citet{Baumgardt21_GCs_membership} estimated the average distance modulus for 162 GCs by compiling a history of literature values (including recent measurements from Gaia EDR3 parallaxes). For well studied clusters, this can be a robust way to estimate the distance modulus. However we noticed for less studied clusters like M9 and M19, there is a large spread in the estimated distance moduli from studies that were often conducted over a decade ago. Therefore, we measure our own distance modulus for each cluster using isochrone fits to recent UBVRI photometry from  \citet{Stetson19_GCS_UBVRI_photometry}. 
  
Both M9 and M19 are located close to the Galactic plane, and thus have large line-of-sight dust extinctions. They also have significant differential reddening across the face of the cluster. To fit isochrones to the cluster colour-magnitude diagram, we need to individually correct the reddening for each star. We implement the differential dust maps from \citet{Alonso_Garcia12_dustmaps}, which provides relative extinctions in the B-V colour ($E(B-V)$) to a zero point value for the cluster. We note that if the zero point extinction is estimated inaccurately, it can introduce a systematic offset in the reddening-corrected photometry. 

We calculate our own zero point extinction for each cluster by using the latest version of the three-dimensional probabilistic map of interstellar dust reddening, Bayestar \citep{Green19_bayestars}. Typically, 2D dust extinction maps --such as SFD \citep{Schlegel1998} or its recalibrated version \citep{Schlafly2011}-- are calculated as the total integrated reddening for a given line of sight. Bayestar is similar to the 2D maps, however it also traces the dust reddening as a function of distance, thus providing a better estimate. The \cite{Alonso_Garcia12_dustmaps} absolute zero points were calculated using the SFD map, which resulted in an overestimation of the line-of-sight extinction values.
By comparing \cite{Alonso_Garcia12_dustmaps} to Bayestar, we can determine new zero points for the relative maps. We query the reddening values at an approximate distance of the cluster in Bayestar, using the same celestial coordinates as the zero point in the \cite{Alonso_Garcia12_dustmaps} map. As the unit of these values is equivalent to the SFD unit in \citet{Schlegel1998}, we convert them to $E(B-V)$ following the suggestions in Section 4.2.1. of \cite{Green2018}. Next, we use a similar approach described by \cite{Alonso_Garcia12_dustmaps}: we take the median of the relative values within each Bayestar pixels, thus bringing down the resolution of the \citet{Alonso_Garcia12_dustmaps} map to the lower resolution of the Bayestar map, and produce one relative extinction value per pixel. The linear relation between the relative and absolute values of the pixels gives us a better approximation for the absolute zero point for the relative reddening maps of the clusters. We find new absolute zero points of $E(B-V)=0.34$ for M9 and $E(B-V)=0.31$ for M19.

\begin{table}
\centering
\footnotesize
\caption{Input parameters used for the two best fitting isochrone models in Figure~\ref{fig:isochrones}. The $(m-M)_0$ parameter is the true distance modulus and was inferred from the isochrone fit to the main sequence turn off. The initial He content ($Y$) is close to the primordial value. The Reimers' mass loss scaling parameter ($\eta_R$) was used to fit to the horizontal branch.}
    \begin{tabular}{lcc}
      \hline 
      Parameter & PARSEC & BaSTI\\
      \hline 
       & \textbf{M9} & \\
      $[\text{M}/\text{H}]$ & -1.270 & -1.398 \\
      $[\alpha/\text{Fe}]$ & 0.0 & +0.4 \\
      Y & 0.2523 & 0.2478 \\
      Age (Gyrs) &  13 & 13 \\
      $\eta_R$ & 0.35 & 0.3 \\
      $(m-M)_{0}$ & 14.70~$\pm~0.13$ & 14.80~$\pm~0.13$ \\
      \hline
       &\textbf{M19} & \\
      $[\text{M}/\text{H}]$ & -1.250 & -1.248 \\
      $[\alpha/\text{Fe}]$ & 0.0 & +0.4 \\
      Y & 0.2522 & 0.2481 \\
      Age (Gyrs) &  12 & 12 \\
      $\eta_R$ & 0.45 & 0.3 \\
      $(m-M)_{0}$ & 14.90~$\pm~0.17$ & 14.95~$\pm~0.17$ \\
      \hline 
    \end{tabular}
\label{tab:isochrone_params}
\end{table}

\begin{figure*}
	\centering
	\includegraphics[width=1.5\columnwidth]{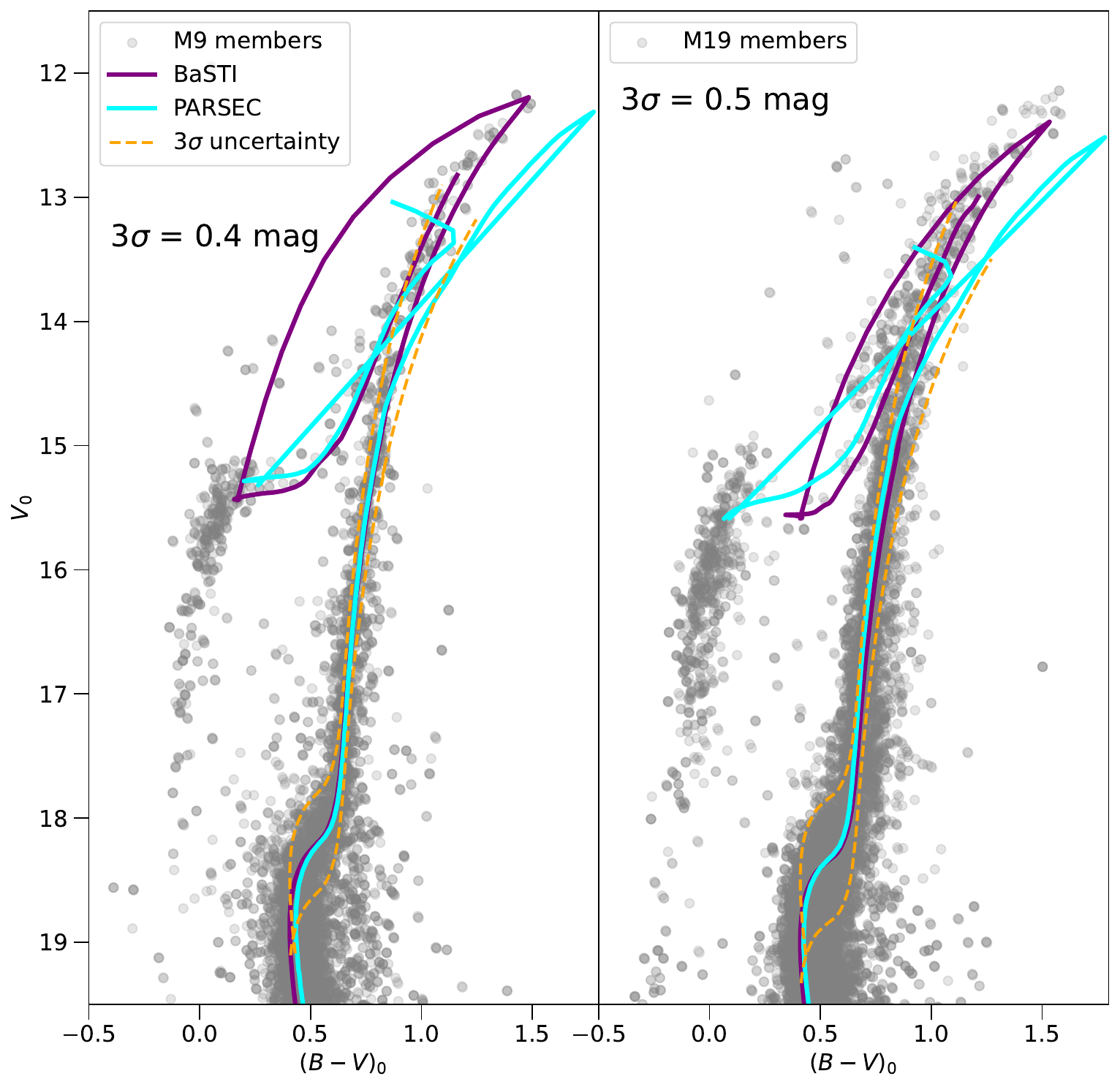}
	\caption{\textbf{Left}: Isochrone fits to dereddenned photometry for a M9 membership sample. UBVRI photometry is from \citet{Stetson19_GCS_UBVRI_photometry} and dust corrections are from \citet{Alonso_Garcia12_dustmaps}. We use two different isochrone models: PARSEC (cyan), and BaSTI (purple). The estimated $3\sigma$ uncertainties on the distance modulus are indicated by the dashed orange lines and amount is written in the text. \textbf{Right}: Same as the left panel but for M19.}
	\label{fig:isochrones}
\end{figure*}

To infer an estimate for the distance modulus, we used two different isochrone databases: PARSEC \citep{Bressan12_PARSEC}, and BaSTI \citep{Pietrinferni21_BaSTI}. We present our isochrone fits to the dereddened photometry in Figure~\ref{fig:isochrones}, and the associated input parameters for these isochrones in Table~\ref{tab:isochrone_params}. The input isochrone parameters were chosen either from previous literature (e.g. ages were from \citealt{Koleva08_M9age,Forbes10_M19age}, and primordial He abundance were from \citealt{Cooke18_premordialHe}) or from the best fits to photometry (e.g. the Reimers' mass loss parameter value was chosen based on the isochrone fit to the horizontal branch, and the metallicity, [M/H], was inferred based on the fit to the main sequence turn off\footnote{Our metallicity estimates are consistent with estimates from Eq.~3 in \citet{Salaris1993}, which use a [Fe/H] and [$\alpha$/Fe] abundance.}). It is known for old, metal-poor stars that they will be enriched in $\alpha$ elements (e.g. O, Mg, Si, S, Ca, Ti; \citealt{Wallerstein1962_alphaenhancement1,Tinsley1979_alphaenhancement2}) by a typical value of [$\alpha$/Fe]=0.4~dex \citep[e.g.][]{GalahDR3}. BaSTI provides the option to compute $\alpha$-enhanced isochrones, however this is not available for PARSEC. To simulate $\alpha$-enhancement in the PARSEC model, we increased the metallicity of the isochrone until it fit the cluster photometry (see Table~\ref{tab:isochrone_params}). We note that M19 has been measured to contain several [Fe/H] and [$\alpha$/H] chemical populations \citep{Johnson_2017_M19,Yong_2016_M19}. However, we aimed to fit the ridge line of the cluster photometry, which we assume represents the main [Fe/H] and [$\alpha$/Fe] populations.   


A systematic uncertainty for the inferred distance modulus was estimated by shifting the isochrones until they were located at either the top or the bottom of the main sequence turn off feature (see orange dashed lines in Fig.~\ref{fig:isochrones}). This was achieved by varying the distance modulus by the same amount in both directions. We assume that fitting to the borders of the photometric data represents the maximum possible values for the inferred distance modulus. We adopt this change in distance modulus as the $3\sigma$ uncertainty, which was measured as 0.4~mag for M9 and 0.5~mag for M19. These adopted uncertainties are similar or slightly larger than previous literature estimates \citep[see list in][]{Baumgardt21_GCs_membership}. We stress that these fits are approximate, and could be overestimating the distance modulus uncertainty. As discussed further in Section~\ref{sec:systematic_uncerts}, the distance modulus is the largest systematic in the seismic mass estimates through the dependence on the luminosity, and hence we were inclined to be cautious in this uncertainty estimate. 

There are also degeneracies in the isochrone input parameters and the inferred distance modulus. \citet{Barker18_M80age5} used a Monte Carlo experiment to investigate the systematic effect to the distance modulus for their isochrone fits to \textit{HST} photometry, by varying the age by 0.5~Gyrs, metallicity by $0.05$~dex and $\alpha$-enhancement by 0.2~dex. They found these variations will result in a change to the distance modulus of $0.05$, $0.05$, and $0.1$, respectively for each input parameter. Hence, their total systematic uncertainty on the distance modulus (when adding these values in quadrature) is $0.12$mag, which is within our inferred uncertainties. 

Taking the average of the inferred values from Table~\ref{tab:isochrone_params}, we estimated a distance modulus of $(m-M)_0 = 14.75\pm0.13$ for M9 and $(m-M)_0 = 14.93\pm0.17$ for M19. Both of these values are larger than the literature value and the distance estimated from the Gaia parallax in \citet{Baumgardt21_GCs_membership}, although they are consistent within 2$\sigma$ uncertainties (as shown in Fig.~\ref{fig:distmod_sys}). This discrepancy is discussed further in Section~\ref{sec:systematic_uncerts}, where we describe why we believe our distance modulus measurements are realistic.  


\subsection{Effective temperatures \& bolometric luminosity}
\label{sec:teffs_dust_lum}

\subsubsection{M9}
\label{sec:teffs_dust_lum_M9}

Using 2MASS $JHK$ \citep{2MASS} and Johnson-Cousins $UBVRI$ magnitudes \citep{Stetson19_GCS_UBVRI_photometry}, we calculated photometric $\Teff$ by implementing the colour-$T_{\text{eff}}$ relation for giant stars from \citet{Gonzales_Hernandez_colour_teff_relation} and a metallicity of [Fe/H]~$ =-1.67\pm0.01\text{(rand)}\pm0.19\text{(sys)}$ \citep{Arellano13_M9FeH}. As shown in \citetalias{Howell22_M4}, differential extinction can have a significant impact on the calculated $\Teff$ values, and hence the asteroseismic masses. Using the differential dust map from \citet{Alonso_Garcia12_dustmaps}, we corrected the V-K colour using Eq.~8 from \citet{Fitzpatrick07_ev_k}. We adopt the same uncertainty of $\pm110~\mathrm{K}$ for the $\Teff$ as \citetalias{Howell_M80}.


We calculated bolometric luminosities ($L$) for each star using the following equation:
\begin{equation}
\label{eq:phot_l}
    \log(L/L_{\odot}) = -0.4\left[V_0-(m-M)_0+BC-M_{\text{bol},\odot}\right]
\end{equation}
and the bolometric correction for giant stars \citep[Eq. 18 from][]{Alonso99_bolometric_correction}. We used the true distance modulus determined in Sec.~\ref{sec:dist_mod}. The $\Teff$ and luminosity estimates for M9 found in Table~\ref{tab:final_results_M9}.

\subsubsection{M19}


We used the same method to calculate the dust extinction corrections and the $\Teff$ values for our M19 sample. This photometric $\Teff$ method requires an estimate for [Fe/H]. As shown by \citet{Johnson_2017_M19} and \citet{Yong_2016_M19}, this cluster has a maximum spread in [Fe/H] of $\sim 1$~dex, significantly larger than the typical metallicity uncertainty of $\sim~0.1$~dex. There are no individual measurements of the Fe abundances for our sample of M19 stars, and we instead adopt the mode metallicity of [Fe/H]~$ =-1.55$~dex from Figure 7 of \citet{Yong_2016_M19} to determine $\Teff$. We conducted a sensitivity test by varying the mode metallicity by the reported intrinsic spread of $\sigma_{\text{intrinsic}} = 0.17$ in the $\Teff$ calculations. This variation in [Fe/H] resulted in offsets in $\Teff$ of $\lesssim~5$~K, which is well within the adopted uncertainty of $110~\mathrm{K}$. 

\begin{figure}
	\centering
	\includegraphics[width=1\columnwidth]{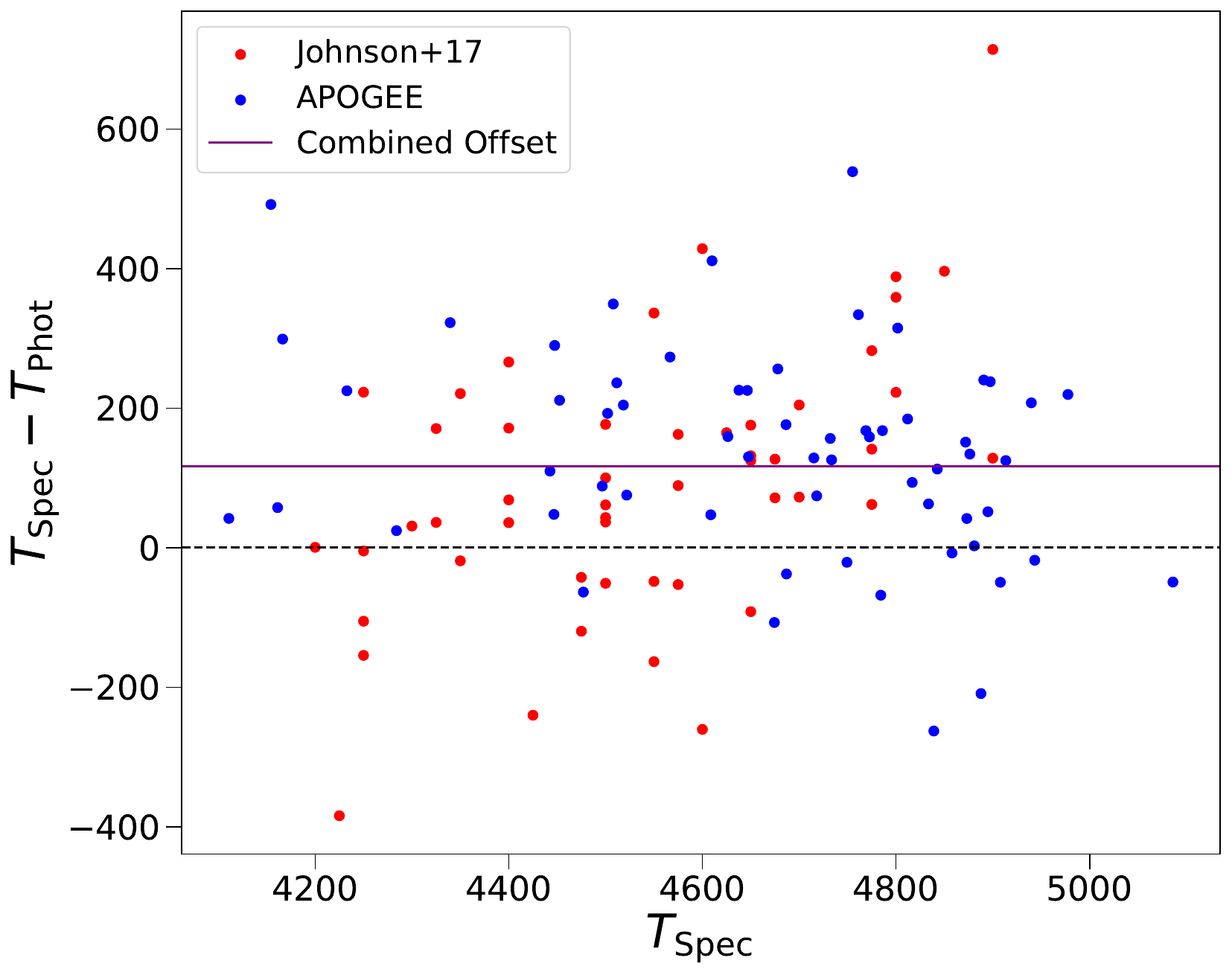}
	\caption{The residuals between our photometric and the spectroscopic $\Teff$ estimates from \citet{Johnson_2017_M19} (red) and the APOGEE DR17 GC catalogue (blue) for M19. By combining the two samples, the mean offset between the two temperature methods was $117\pm16$~K (purple line).}
	\label{fig:Teff_offsets}
\end{figure}

Ideally, we need dust-independent spectroscopic $\Teff$ to remove the systematic uncertainties introduced by the extinction corrections (see discussion in Sec.~\ref{sec:dist_mod}). The APOGEE DR17 GC catalogue \citep{Abdurrouf2022_APOGEEDR17,Schiavon24_APOGEEGCs} measured the spectroscopic temperatures for nine stars in our seismic sample\footnote{In Table~\ref{tab:final_results_M19} we have identified the stars with spectroscopic $\Teff$ measurements with asterisks.}. Thus, we adopted the spectroscopic values for these stars. For the rest of the sample, we followed a similar method as \citetalias{Howell22_M4} where our photometric temperatures were scaled by an offset value calculated as the difference between spectroscopic and photometric methods. To determine the offset, we used the spectroscopic $\Teff$ estimates for red giants from \citet{Johnson_2017_M19} and the APOGEE SDSS-IV GC catalogue, and calculated the photometric $\Teff$ using the same method as Section~\ref{sec:teffs_dust_lum_M9}. Figure~\ref{fig:Teff_offsets} shows the residuals of the spectroscopic and photometric temperatures for the \citet{Johnson_2017_M19} and the APOGEE samples. We calculated an offset of $117\pm16$~K, which was estimated as the mean difference of the combined samples. This offset was then added to the photometric temperatures for stars with no spectroscopic measurements.

Finally, the bolometric luminosities for our M19 sample were calculated in the same way as M9, adopting the true distance modulus found in Sec.~\ref{sec:dist_mod}. The $\Teff$ and luminosity estimates for M19 can be found in Table~\ref{tab:final_results_M19}.

\section{Mass Results \& Discussion}
\label{sec:mass_results}
\subsection{Measuring the seismic masses and integrated mass loss}
\label{sec:seismic_masses_massloss}
\begin{table*}
\centering
\footnotesize
\caption{Average masses for each evolutionary phase ($\overline{M}$) and integrated mass loss ($\Delta M_{\text{RGB}-\text{EAGB}}$) for M9 and M19. The random (rand) uncertainty for the average masses is calculated as the standard error on the mean, which are added in quadrature for the uncertainty on the integrated mass loss. In Sec.~\ref{sec:systematic_uncerts}, we detail how the systematic (sys) uncertainties are calculated.}
    \begin{tabular}{lccc}
      \hline 
      Cluster & $\overline{M}_{\text{RGB}}$ & $\overline{M}_{\text{EAGB}}$ & $\Delta M_{\text{RGB}-\text{EAGB}}$ \\
      \hline 
       M9 & $0.78\pm0.01$(rand){\raisebox{0.5ex}{\tiny$\substack{+0.13 \\ -0.15}$}}(sys) & $0.62\pm0.02$(rand){\raisebox{0.5ex}{\tiny$\substack{+0.10 \\ -0.12}$}}(sys) & $0.16\pm0.02$(rand){\raisebox{0.5ex}{\tiny$\substack{+0.03 \\ -0.03}$}}(sys)  \\
       M19 & $0.83\pm0.01$(rand){\raisebox{0.5ex}{\tiny$\substack{+0.22 \\ -0.17}$}}(sys) & $0.50\pm0.02$(rand){\raisebox{0.5ex}{\tiny$\substack{+0.13 \\ -0.10}$}}(sys) & $0.33\pm0.03$(rand){\raisebox{0.5ex}{\tiny$\substack{+0.09 \\ -0.07}$}}(sys) \\
      \hline 
    \end{tabular}
\label{tab:average_masses}
\end{table*}

\citetalias{Howell22_M4} and \citetalias{Howell_M80} showed that accurate asteroseismic masses could be estimated without a $\Delta\nu$ measurement. They used estimates for $\numax$, $\Teff$, and $L$ to calculate masses with the following relation\footnote{We adopted the following solar reference values: $\nu_{\text{max},\odot} = 3090\pm30~\mu\mathrm{Hz}$ \citep{Huber11_solar_syd_values} and $\Teffsun = 5772\pm0.8~\mathrm{K}$ \citep{Mamajek15_B3}.}:
\begin{align}
    \label{eq:mass_relation3}
    &\left(\frac{M}{\msun}\right)\simeq\left(\frac{\nu_{\text{max}}}{\nu_{\text{max},\odot}}\right)\left(\frac{L}{\Lsun}\right)\left(\frac{T_{\text{eff}}}{\Teffsun}\right)^{-7/2}
\end{align}
We adopted this same method to calculate individual stellar masses for our M9 and M19 red giants. We also estimated a seismic $\log{(g)}$ for each star using:
\begin{equation}
    g/g_{\odot} \simeq \left(\frac{\nu_{\text{max}}}{\nu_{\text{max},\odot}}\right)\left(\frac{T_{\text{eff}}}{\Teffsun}\right)^{-1/2}
\end{equation}
The final seismic mass and $\log{(g)}$ estimates are provided in Table~\ref{tab:final_results_M9} for M9 and Table~\ref{tab:final_results_M19} for M19.

\begin{figure*}
	\centering
	\includegraphics[width=2\columnwidth]{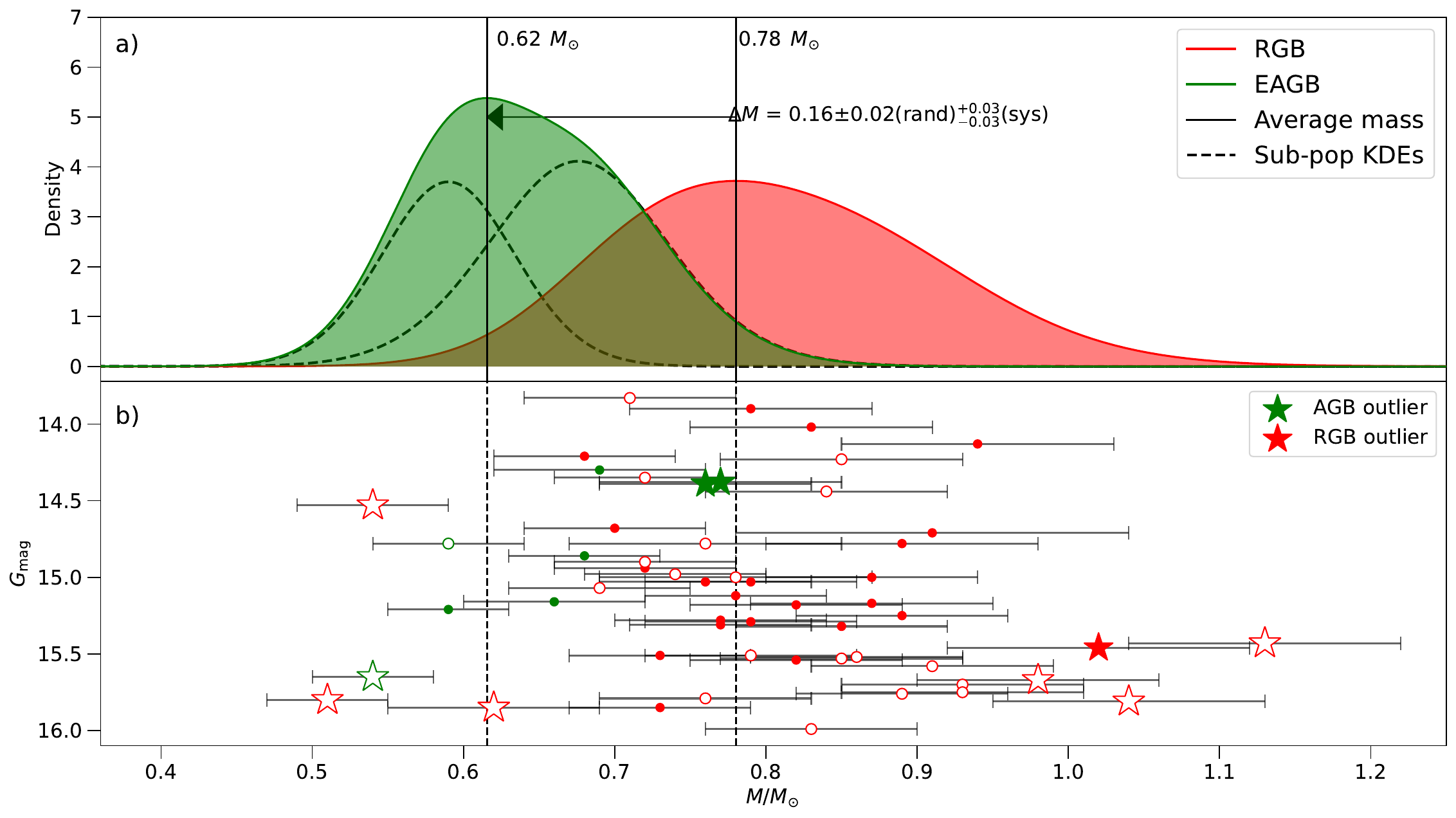}
	\caption{\textbf{a)} The mass distribution for our RGB (red) and EAGB (green) samples for M9, calculated as KDE functions. The measured average masses (modes) for the RGB and EAGB evolutionary phase are shown by black vertical lines and the values are annotated. The black arrow indicates the direction of the integrated mass loss, and the amount of mass loss is annotated. We note that the area for each KDE is normalised to one, and hence the heights of the individual mass distributions are not representative of sample sizes. We have also decomposed the EAGB mass distribution into hypothesised sub-population KDEs (black dashed functions; see Sec.~\ref{sec:multi_pops} for further details). \textbf{b)} Individual masses plotted against \textit{Gaia} DR3 magnitudes. Stars with a `MD' quality flag are identified by open circles. The mass outliers are illustrated by star symbols (refer to Sec.~\ref{sec:seismic_masses_massloss} and \ref{sec:mass_outliers} for further details).}
	\label{fig:M9_KDE}
\end{figure*}

\begin{figure*}
	\centering
	\includegraphics[width=2\columnwidth]{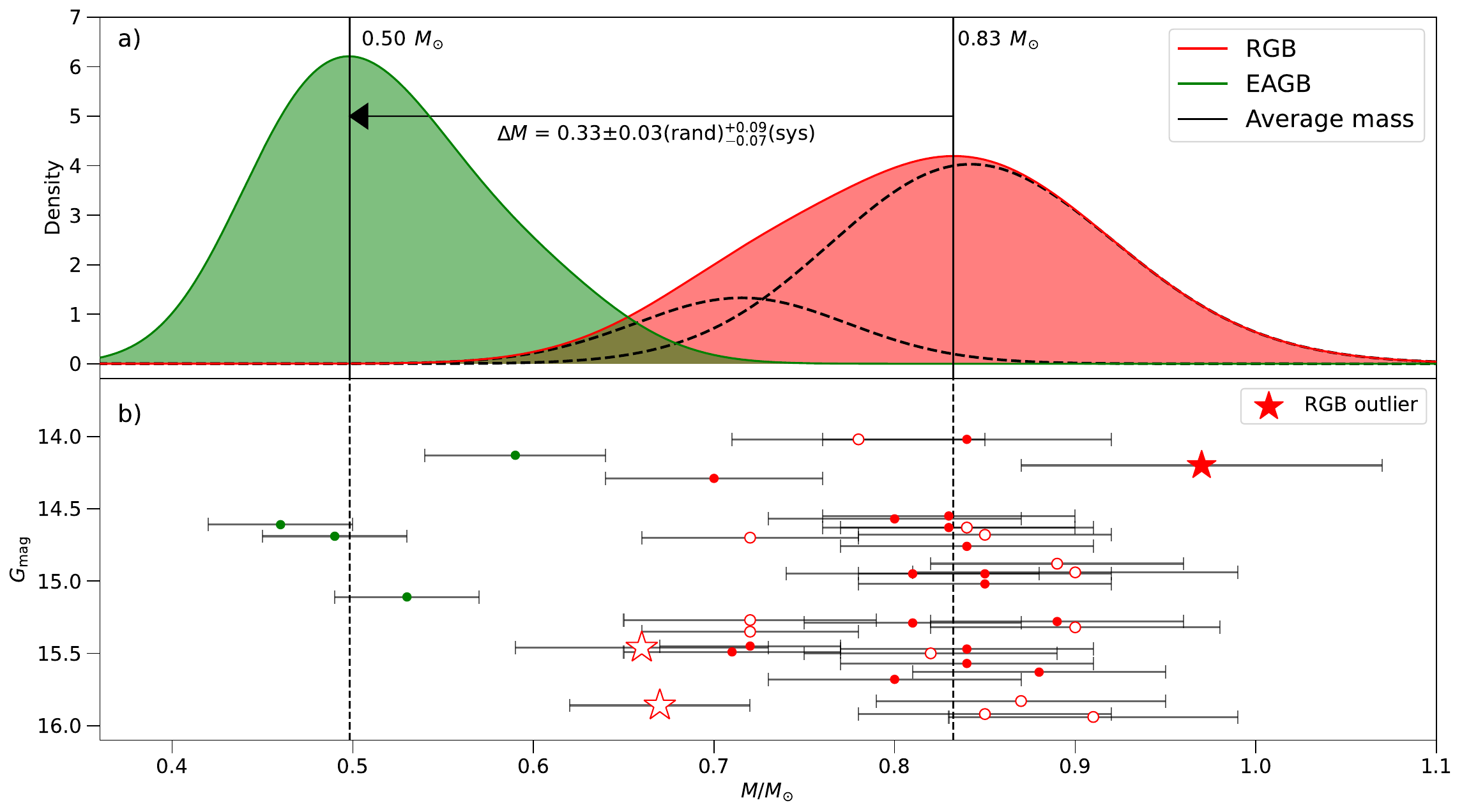}
	\caption{Same as Figure~\ref{fig:M9_KDE} but for M19. We have decomposed the RGB mass distribution into hypothesised sub-population KDEs (black dashed functions; see Sec.~\ref{sec:multipops_M19} for further details).}
	\label{fig:M19_KDE}
\end{figure*}

Figures~\ref{fig:M9_KDE} (M9) and \ref{fig:M19_KDE} (M19) show the seismic mass distributions separated into evolutionary phases. The distributions are represented as kernel density estimation (KDE) functions (see \citetalias{Howell22_M4} and \citetalias{Howell_M80} for details), which take into account the individual random uncertainties. To not introduce a bias to the distributions, for each KDE we have excluded stars with outlying masses identified through an iterative sigma-clipping process; masses that were not within $2\sigma$ of the mean mass for each evolutionary phase were removed (where $\sigma$ was calculated as the standard error on the mean). This process was completed three times. We discuss the potential explanations for the outlying masses in Section~\ref{sec:mass_outliers}.

\begin{figure*}
	\centering
	\includegraphics[width=2\columnwidth]{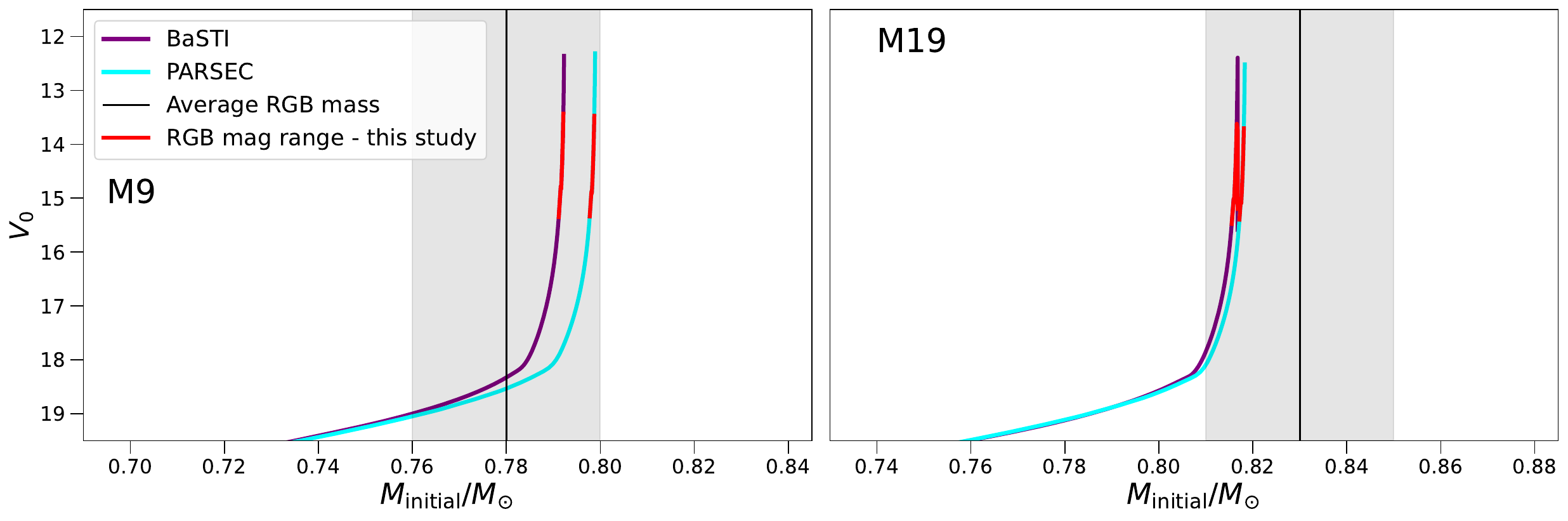}
	\caption{Comparison of the initial masses from the BaSTI and PARSEC isochrones to the measured average RGB mass (black line) and $2\sigma$ random uncertainty (grey shaded region). The magnitude range for our RGB sample in the dereddened V-band is indicated in red. }
	\label{fig:isochrone_masses}
\end{figure*}

Using these KDEs, we measured the average mass for each evolutionary phase as the mode of the distribution. This value represents the most common mass for the sample, and is provided in Table~\ref{tab:average_masses}. As a test, we checked whether the mass outliers removed via sigma clipping would alter our average mass estimates. We found that including the mass outliers only introduced small tails at the ends of our KDEs. Thus, they had insignificant effects on the mode measurements from each mass distribution. We adopt the standard error on the mean as the uncertainty for the average masses. We compare our average mass estimates to the initial masses from isochrones for our RGB magnitude range in Figure~\ref{fig:isochrone_masses}. Our average masses for the RGB in both clusters are consistent within 2$\sigma$ uncertainties to the isochrone initial mass. This result is similar to \citetalias{Howell22_M4} and \citetalias{Howell_M80}, further confirming that reliable seismic masses can be measured for low-metallicity red giants using a $\Delta\nu$-independent relation. 

The average mass for our EAGB sample in M19 of $0.50\pm0.02~\msun$ is lower than typically modelled for this evolutionary phase, and is similar to GC white dwarf masses ($0.50-0.55~\msun$; \citealt{Kalirai09_WD_masses}). This was also seen in M4 \citepalias{Howell22_M4}, where they measured an average mass of $0.54\pm0.01~\msun$ for their EAGB sample. They suggested that the lower masses indicate that these stars have thin envelopes, which will be lost quickly. Due to this, they will skip the thermally pulsing AGB stage and evolve to become a white dwarf, similar to AGB-manqu\'e stars \citep{Dorman1995_agb_manque}. Alternatively, this result may be due to the small sample size of four stars in both M4 and M19, or the seismic mass relation not being robust for these stars. Hence, the average mass calculation may not represent the true mass of the EAGB sample for those clusters. We would need a larger sample of seismically measured EAGB stars to confirm this. 

We estimated the integrated mass loss as the difference between the average masses of various evolutionary phases. Similar to \citetalias{Howell_M80}, M9 and M19 were too faint to measure seismic masses of the HB stars. Hence, we were not able to directly measure the RGB mass loss. Instead, we can calculate the mass difference between the RGB and EAGB ($\Delta M_{\text{RGB}-\text{EAGB}}$; see Table~\ref{tab:average_masses} for values). This is illustrated by the arrows in Figures~\ref{fig:M9_KDE} and \ref{fig:M19_KDE}. We compare our measured mass loss estimates to other clusters in Section~\ref{sec:massloss_metallicity}. 


\subsection{Systematic uncertainties on the masses: distance modulus}
\label{sec:systematic_uncerts}
\begin{figure*}
	\centering
	\includegraphics[width=2\columnwidth]{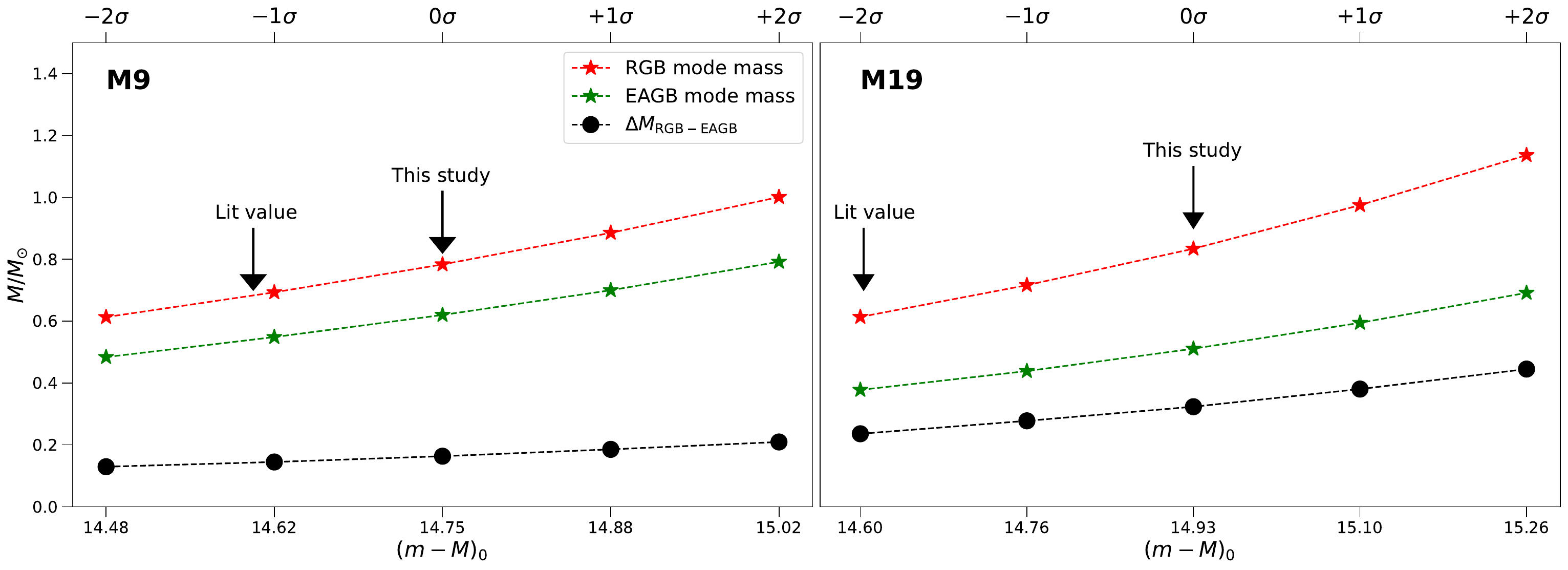}
	\caption{The trend in average mass for RGB (red) and EAGB (green), and mass difference between the averages (black) when varying the distance modulus by a maximum of $\pm2\sigma$ (as indicated on the top y-axis) for M9 (left) and M19 (right). The central value is the inferred distance modulus found in Sec~\ref{sec:dist_mod}\mh{\sout{)}}. The left arrows signify the average distance modulus estimated from the literature search in \citet{Baumgardt21_GCs_membership}. }
	\label{fig:distmod_sys}
\end{figure*}

In Section~\ref{sec:dist_mod}, we estimated the distance modulus of both clusters from isochrones and briefly stated that this parameter has a large systematic effect on the seismic masses through the dependence on the luminosity. We demonstrate this in Figure~\ref{fig:distmod_sys}, where we show the trend in the average masses and the mass loss when the distance modulus is varied by up to $\pm2\sigma$ (see Table~\ref{tab:isochrone_params} for systematic uncertainty on the distance modulus). The change in mass increases monotonically with distance modulus. Hence, we see the greatest change in mass for the average RGB mass at larger distance moduli. We used Figure~\ref{fig:distmod_sys} to estimate the systematic uncertainty for the mass estimates, by adopting the change in mass at $\pm1\sigma$ (see Table~\ref{tab:average_masses}). We found that the systematic uncertainties for the mass loss estimates are the same order of magnitude as the random uncertainty. In contrast, the systematic uncertainties for the average masses are an order of magnitude larger. This implies that even if there is a large systematic dependency for the average masses, this is not the case for mass loss, i.e. the integrated mass loss result is not sensitive to the uncertainty on the distance modulus. 

We demonstrate this further by comparing our mass estimates to the mass estimates calculated with the literature value for the distance modulus. As indicated with the left arrows in Figure~\ref{fig:distmod_sys}, the literature distance moduli \citep{Baumgardt21_GCs_membership} for both clusters are less than our inferred values, however they are consistent within $2\sigma$ uncertainties. Again, there is a significant change in the average mass of the evolutionary phases when using the literature distance modulus (between $0.08-0.21~\msun$), although the change in mass loss is smaller: $0.02~\msun$ for M9 and $0.10~\msun$ for M19. 

From our systematic tests in Figure~\ref{fig:distmod_sys}, we are also confident in our measurement for the distance modulus. By varying the distance modulus by 1-2~$\sigma$, the average masses would no longer be consistent with the isochrone models and would move into mass regimes which are unrealistic:
\begin{enumerate}
    \item When varying the distance modulus beyond $ +1\sigma$, the average RGB masses would be larger than the expected GC mass from isochrones. This would lead to the conclusion that the clusters are either enriched in more metals or are much younger than previously measured for these GCs (i.e. less than 10~Gyrs). Both of these scenarios are unlikely (also note that the metallicity is well known for these clusters).
    \item When varying the distance modulus beyond $-1\sigma$, the average RGB masses are smaller than expected. In this case the RGB stars would be older than the age of the Universe, which immediately excludes the lower distance modulus values. Additionally, the EAGB average masses are smaller than the expected core mass ($\sim0.48~\msun$; \citealt{Dorman1995_agb_manque}).  
\end{enumerate}
This supports our measurements of the distance modulus in Section~\ref{sec:dist_mod} for both clusters, given that our original RGB average masses are consistent with isochrones (see Fig.~\ref{fig:isochrone_masses}). 

To reiterate, the choice of distance modulus does have a significant systematic effect on the calculated average masses, however this effect is minimised for the calculated mass loss (Fig.~\ref{fig:distmod_sys}). As such, asteroseismology is an excellent tool to measure the integrated mass loss for red giants in GCs despite the uncertainties in the distance modulus.

\subsection{Multiple populations: bimodal mass distribution?}
\label{sec:multi_pops}

\subsubsection{M9}
It is suspected that every GC contains multiple sub-populations that vary in light chemical elements. Due to the variation in He, the stars in each sub-population have a different initial mass \citep[e.g.][]{MacLean18_chloes_paper,Jang19_2pops_models,Tailo20_massloss_difference_multipops}. From models, the mass difference between the sub-populations is calculated to be a small amount ($\sim 10^{-2}~\msun$). However, for some clusters the difference can be as large as $0.17~\msun$ \citep{Tailo20_massloss_difference_multipops}. This mass difference is currently only theoretical; it has not been definitively confirmed by direct mass measurements. 

The most direct evidence thus far of a mass difference between sub-populations was the detection of a bimodal mass distribution in the EAGB sample in \citetalias{Howell_M80}. Furthermore, the peaks of their bimodal distribution (assumed to correspond to the sub-population masses) matched model estimates for the sub-populations of this cluster \citep{Tailo20_massloss_difference_multipops}. This could be the first direct detection of a mass difference between the sub-populations. However, measurements of light element abundances are required to confirm that the bimodality is correlated to chemically-defined sub-population classification. Similarly, for our study we cannot conclusively find evidence of a mass difference because we do not have spectroscopic measurements of our total samples. However, here we attempt to detect a mass difference between the sub-populations by analysing our KDE mass distributions.

For our analysis, we assume that there are two main sub-populations, and adopt the following naming convention:
\begin{enumerate}
    \item Sub-population 1 (SP1): thought to be the first generation of stars with a chemical composition similar to Galactic field stars at the same [Fe/H].
    \item Sub-population 2 (SP2): the supposed second generation of stars in the cluster, which are enriched in certain light chemical elements, such as He, N and Na. Due to their larger He abundances, they are expected to have lower masses compared to SP1 stars (assuming no significant difference in age). 
\end{enumerate}
By employing this two sub-population prescription, we will search for a bimodal mass distribution.   


The RGB KDE (red) in Figure~\ref{fig:M9_KDE} does not show any evidence of a bimodal signal. This is further supported by looking at the individual mass points in the bottom panel, which contains a large scatter and no clear bimodal pattern. However, the width of the KDE is quite large with tails to both high and low masses. Similar to \citetalias{Howell22_M4} and \citetalias{Howell_M80}, we speculate that there could be a small hidden mass difference (most likely of the order $10^{-2}~\msun$) in the KDE. Spectroscopy could help here by differentiating our RGB sample into each sub-population with chemical abundances.   

In comparison, the EAGB KDE (green) in Figure~\ref{fig:M9_KDE} shows a possible signature of a bimodality as seen by the shoulder at a mass of $\sim0.68~\msun$, and the bimodal pattern in the individual mass points. As a test, we split our EAGB sample into two groups: stars with masses $<0.65~\msun$ and $>0.65~\msun$. We then calculated new KDEs for each sub-group (shown by dashed black functions in Fig.~\ref{fig:M9_KDE}). Using the mean of these KDEs, we estimated sub-population masses of $0.67\pm0.03~\msun$ (SP1) and $0.59\pm0.03~\msun$ (SP2), where the uncertainty is calculated as the standard error on the mean. This corresponds to a mass difference between the sub-population of $0.09\pm0.04~\msun$. However, we stress that this is a small sample, with a total of 5 EAGB stars and as such we cannot conclusively claim that this is a detection. We again need spectroscopic measurements of light elements to classify the sub-populations, and hence confirm if this potential signature is a detection of a sub-population mass difference.

It is more likely that we can detect a bimodal signature in the EAGB and not the RGB. This is because SP2 stars are suspected to have a larger RGB mass loss compared to SP1 stars \citep[e.g.][]{Rood73_2ndparam,Dorman93_2ndparam2,Dcruz96_2ndparam3,Ferraro98_2ndparam4,Whiteney98_2ndparam5,Catelan00_2ndparam6, Tailo19_M4_multipop_massloss}. This will increase the mass difference between sub-populations in the distributions of evolutionary phases after the RGB. Although the concept of enhanced mass loss for SP2 has been modelled \citep{Tailo20_massloss_difference_multipops} and shown in tentative detections \citepalias{Howell_M80}, there is no definitive explanation for why SP2 will lose more mass on the RGB. One theory proposed by \citet{Tailo20_massloss_difference_multipops}, is that SP2 stars have faster core rotation at the end of the main sequence, which delays the ignition of the core He flash and increases the RGB lifetime. Hence, SP2 stars will have more time to lose mass and as such will have a larger integrated mass loss. Direct concrete measurements of the difference in sub-population mass loss are still needed to substantiate this theory.  



\subsubsection{M19}
\label{sec:multipops_M19}
M19 is a Type II cluster, which is characterised as having non-homogeneous heavy element distributions (e.g. in Fe and s-process elements). In \citet{Johnson15_M19, Johnson_2017_M19} and \citet{Yong_2016_M19}, three Fe populations were identified; `metal-poor' with ([Fe/H]~$\leq -1.65$), `metal-intermediate' ($-1.65<$~[Fe/H]~$\leq -1.35$), and a metal-rich tail ([Fe/H]~$\geq-1.35$). For a given age, we would expect to see a mass difference between these Fe populations, because of the dependence of metallicity in the initial mass of a star. 

The presence of these Fe populations complicates the multiple population picture; each individual Fe population contains its own set of sub-populations. Trying to disentangle the mass signatures of six or more sub-populations in Figure~\ref{fig:M19_KDE} is impossible without at least classifying the sample into the Fe populations. Furthermore, the mass difference between the Fe populations would be in addition to the signatures of a light element sub-population mass difference. For simplicity, we assume that any multi-modal signature in our mass distributions would be due to the difference in Fe abundances in the cluster.

Initially, we investigated what the expected mass difference for the RGB sample would be between the three Fe populations using isochrones. For the isochrones in Figure~\ref{fig:isochrone_masses}, a metallicity consistent with the metal-intermediate population was used, and from that we found a RGB isochrone mass of $M = 0.82~\msun$. Keeping the other parameters the same, we created new isochrones to represent the metal-poor and metal-rich populations by varying the metallicity by [M/H]~$=\pm0.2$. We measured a mass difference of the order of $10^{-2}~\msun$ between the Fe populations. This is similar to the expected mass difference of the light-element sub-populations. The metal-poor and metal-intermediate populations contribute to the majority of the total number of stars in this cluster (approximately 48\% each). Because the metal-rich tail sample make up only a small proportion of the total sample, we assume that we will not detect a mass signature for this population, and thus expect to only observe a bimodal mass distribution. Although, we note that this mass difference may be too small to detect with our mass uncertainties.

There is no clear bimodal signal in the RGB or EAGB mass distributions in Figure~\ref{fig:M19_KDE}. However, the RGB KDE contains an extended tail to lower masses. Similar to M9, as a test we break our RGB sample into two groups, and create new KDEs (black dashed lines in Fig.~\ref{fig:M19_KDE}), separating the RGB sample at a mass of $0.75~\msun$. We chose this mass to split the samples due to an observed natural division in the individual masses in the bottom panel of Figure~\ref{fig:M19_KDE}. From the new KDEs, we measured a mass difference between the Fe populations of $0.13\pm0.03~\msun$ for RGB\footnote{Similarly to the M9 analysis, the uncertainty on the mass for each population was calculated as the standard error on the mean, and then added in quadrature to determine the mass difference uncertainty.}. This is significantly larger than the mass difference we found in our isochrone tests. We suggest that this could be due to the combination of the mass difference between the Fe populations and the light-element sub-populations, resulting in a larger mass difference. However, we cannot prove this within the scope of this paper and again would require measurements of the chemical abundances. We do have metallicity measurements from APOGEE \citep{Abdurrouf2022_APOGEEDR17,Schiavon24_APOGEEGCs} for nine RGB stars, however there are only four measurements of Na. As such, we are unable to do in-depth analysis on this sub-sample. 

Intriguingly, our inferred measurement for the mass difference between the Fe population KDEs for the RGB is larger than for the EAGB sample. This is opposite to what we found for M9. We speculate that this is because there is a trend in the amount of mass loss at different [Fe/H] abundances; i.e. the metal-intermediate population would have a larger mass loss compared to the metal-poor population. However as already mentioned, without Fe measurements we cannot investigate this further.


\subsection{Preliminary mass loss-metallicity trend}
\label{sec:massloss_metallicity}
\begin{figure}
	\centering
	\includegraphics[width=1\columnwidth]{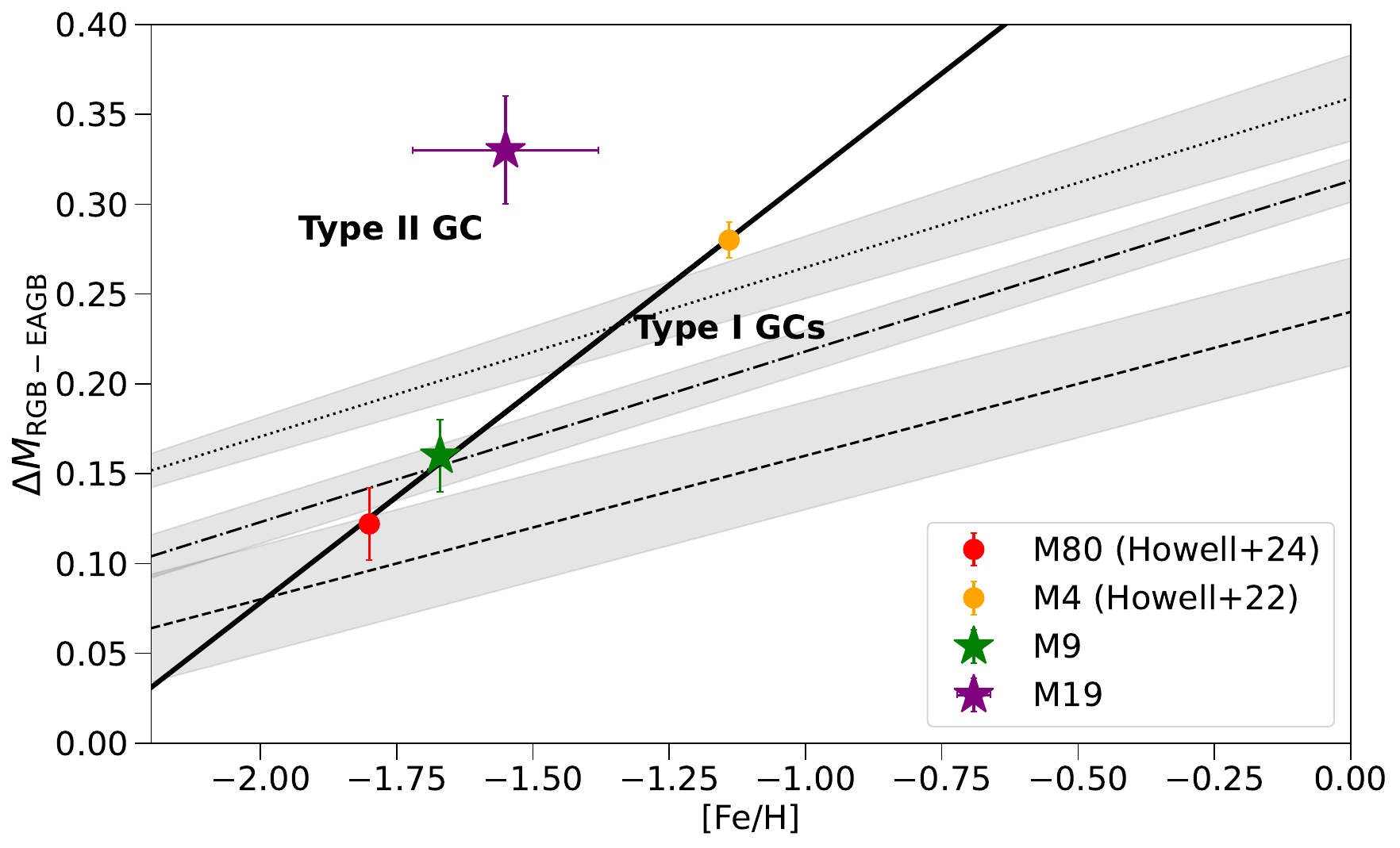}
	\caption{Measurements of the integrated mass loss ($\Delta M_{\text{RGB-EAGB}}$) between the RGB and EAGB evolutionary phases for four GCs: M9, M19, M4 \citepalias{Howell22_M4} and M80 \citepalias{Howell_M80}. M19 has a range in [Fe/H], which is represented by the error bar. We include three RGB mass loss-metallicity trends and their uncertainty (grey shaded) from \citealt{Gratton10_masslossGCs} (black dotted), \citealt{Origlia14_masslossGCs} (black dashed), and \citealt{Tailo20_massloss_difference_multipops} (black dashdot). From the Type I clusters (i.e. excluding M19) we derive a preliminary mass loss-metallicity trend (thick black solid).}
	\label{fig:massloss_metallicity}
\end{figure}

It has been demonstrated with models \citep{Gratton10_masslossGCs,Tailo20_massloss_difference_multipops} and circumstellar dust measurements \citep{Origlia14_masslossGCs} of GCs that there is a relationship between RGB mass loss and metallicity. These studies showed that lower metallicity stars will lose less mass compared to their higher-metallicity counterparts. Asteroseismic mass loss measurements are also in agreement with this concept, where \citetalias{Howell_M80} found that their estimated integrated mass loss for the metal-poor cluster M80 was less than the mass loss for the more metal-enriched cluster M4 \citepalias{Howell22_M4}. Figure~\ref{fig:massloss_metallicity} shows the extension to Figure 11\footnote{We note that \citetalias{Howell_M80} plotted the SP1 mass loss. Because we aren't able to definitively separate our mass distributions into sub-populations for the GCs in this study, we only plot our mass loss estimates calculated from the peaks of our KDEs. Hence, our measurements could be the combination of the sub-population mass loss.} in \citetalias{Howell_M80}, where we have added the asteroseismically calculated mass loss for M9 and M19. For simplicity, we only plot the $\Delta M_{\text{RGB-EAGB}}$ estimates, excluding any previous measurements of RGB mass loss (i.e. $\Delta M_{\text{HB-EAGB}}$). As a guide, we include the three RGB mass loss-metallicity trends from the previous non-asteroseismic studies \citep{Gratton10_masslossGCs, Origlia14_masslossGCs, Tailo20_massloss_difference_multipops}.

In Section~\ref{sec:multi_pops}, we presented evidence of a weak detection of dual mass signature in the RGB, which could correspond to different Fe populations in M19. However, we cannot split our sample into the Fe abundances, and as such cannot measure accurate mass loss for each population. To remain agnostic, we only consider the mass difference between the average masses for the entire RGB and EAGB sample (Table~\ref{tab:average_masses}). In Figure~\ref{fig:massloss_metallicity}, we represent the metallicity of this cluster as the average [Fe/H] from \citet{Yong_2016_M19}, with the intrinsic spread as the uncertainty. We find that the mass loss estimates for M19 are larger than estimates from Type I GCs. Similar to the enhanced mass loss for SP2 stars theory in \citet{Tailo20_massloss_difference_multipops} (see Sec.~\ref{sec:multi_pops} for further details), we speculate that the formation process of Type II clusters could lead to longer lifetimes on the RGB, which results in more matter being lost compared to Type I clusters. Alternatively, there could be a larger mass loss on the HB compared to Type I clusters. In summary, Figure~\ref{fig:massloss_metallicity} suggests that there is another process that is causing stars in Type II clusters to eject more mass compared to Type I clusters at a fixed metallicity. 

In contrast, our estimated integrated mass loss for M9 is consistent with the previous asteroseismic calculations for the Type I clusters: M4 and M80. Excluding the mass loss estimates for M19, we calculated our own mass loss-metallicity trend that can be tentatively used to infer a mass loss value for Type I GCs with [Fe/H]$<0$: 
\begin{equation}
    \Delta M_{\text{RGB-EAGB}} = 0.24~\text{[Fe/H]}+0.55
\end{equation}
This relation is illustrated by the thick black line in Figure~\ref{fig:massloss_metallicity}. This is the first mass loss-metallicity relation that uses direct model-independent mass measurements. \textit{However, we stress that this trend is preliminary as it is derived from only three mass loss measurements}.


Furthermore, our mass loss-metallicity relation has a steeper gradient compared to the background trends (grey) in Figure~\ref{fig:massloss_metallicity}. Our trend is calculated from the mass difference between the RGB and EAGB, whereas the background trends are calculated just as the RGB mass loss. This implies that there could be significant mass loss on the HB for more metal-enriched clusters. Indeed this is what \citetalias{Howell22_M4} found for their asteroseismic HB sample in M4, where they measured a HB mass loss of $0.12\pm0.01~\msun$. In contrast, \citetalias{Howell_M80} inferred that there was insignificant mass loss on the HB for the more metal-poor GC M80 when comparing their EAGB masses to HB models. To derive a more precise and accurate mass loss-metallicity trend, we need to increase our sample size by measuring asteroseismic masses for stars in more GCs. This requires more time-series photometric data of GCs with longer base-lines than the \textit{K2} campaigns (to increase signal-to-noise and frequency resolution for faint targets such as HB stars).

\subsection{Mass outliers}
\label{sec:mass_outliers}

\begin{figure}
	\centering
	\includegraphics[width=1\columnwidth]{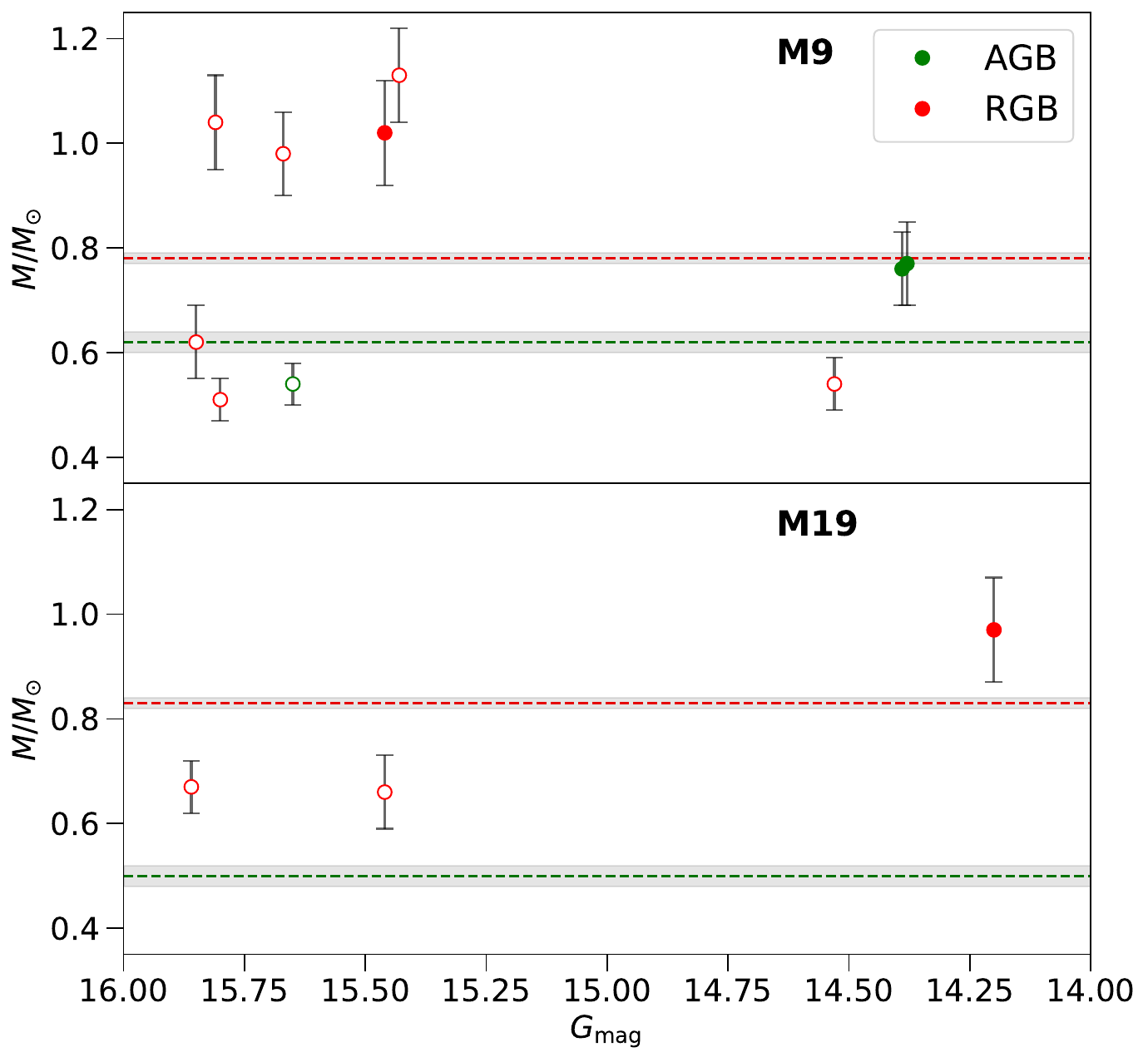}
	\caption{\textbf{Top}: Stars with outlying masses compared to the measured average masses for each evolutionary phase (horizontal lines) for M9. The random uncertainties are demonstrated in grey for both the averages and individual mass points. The quality flags assigned to each star are represented by open circles for marginal detections or closed circles for true detections. On the x-axis we plot the \textit{Gaia} DR3 magnitudes, where evolution up a branch for both the RGB or EAGB occurs from left to right. \textbf{Bottom:} Same but for M19. }
	\label{fig:mass_outliers}
\end{figure}

To calculate the average mass for each evolutionary phase, we removed stars with outlying masses through an iterative sigma clipping process (see Sec.~\ref{sec:seismic_masses_massloss} and bold text star IDs in Tables~\ref{tab:final_results_M9} \& \ref{tab:final_results_M19}). Figure~\ref{fig:mass_outliers} demonstrates the discrepancy between the outlying masses and their corresponding evolutionary phase average mass. We note that our sigma clipping method only considered the uncertainty on the average mass, and didn't account for the larger individual mass uncertainties. This ensured that these outlying masses did not produce tails in the KDEs in Figures~\ref{fig:M9_KDE} \& \ref{fig:M19_KDE}, even if they had large uncertainties that would make them consistent with the average. Because the average evolutionary mass is estimated as the peak of the KDEs, we showed that excluding these outliers does not have a significant effect on this measurement. 

Mass outliers were also detected in \citetalias{Howell22_M4} and \citetalias{Howell_M80} where it was hypothesised that they could either be: stars that have atypical evolution compared to the rest of the GC sample (e.g. mergers or mass stripping events), or incorrect classification of the evolutionary phase. There is also a chance that we have incorrectly measured one of the parameters that are used in the seismic mass relation. We will explore these options here.

As discussed in detail in \citetalias{Howell22_M4} and \citetalias{Howell_M80}, measurements of asteroseismic masses have been used to identify products of non-standard evolutionary events in open clusters such as blue straggler stars (e.g. \citealt{Brogaard12_seismo_eBSs_obs1,Brogaard21_non_standard_evol_seismo}; \citealt{Corsaro12_seismo_eBSs_obs2}; \citealt{Handberg17_OC_study}; \citealt{Leiner19_bluelurkers}; \citealt{Reyes2024}) and possible stripped He-burning field stars \citep{Li22_nature}. Non-standard evolutionary events could explain why our outlying stars do not have masses consistent with the average. For example, at the faint end of Figure~\ref{fig:mass_outliers} for M9 there is a sample of RGB stars with masses $M\gtrsim 1~\msun$. Although they are not large enough to be the product of two $\sim0.8~\msun$ stars merging (typically $1.2-1.6~\msun$; \citealt{Tian06_eBSS_models1,Sills09_eBSS_models2}), they could have accreted some mass through Roche-lobe overflow in a binary system. Similarly, the under-massive stars could have donated mass to another star in a stripping event. However, we cannot confirm whether our outliers are the products of a merger or stripping event with only asteroseismic masses, and there are no extensive binary surveys of these clusters. There are four known eclipsing binaries in M9 \citep{Clement01_VariablesM19, Arellano13_M9FeH}, although they aren't within our sample. 

\begin{figure}
	\centering
	\includegraphics[width=1\columnwidth]{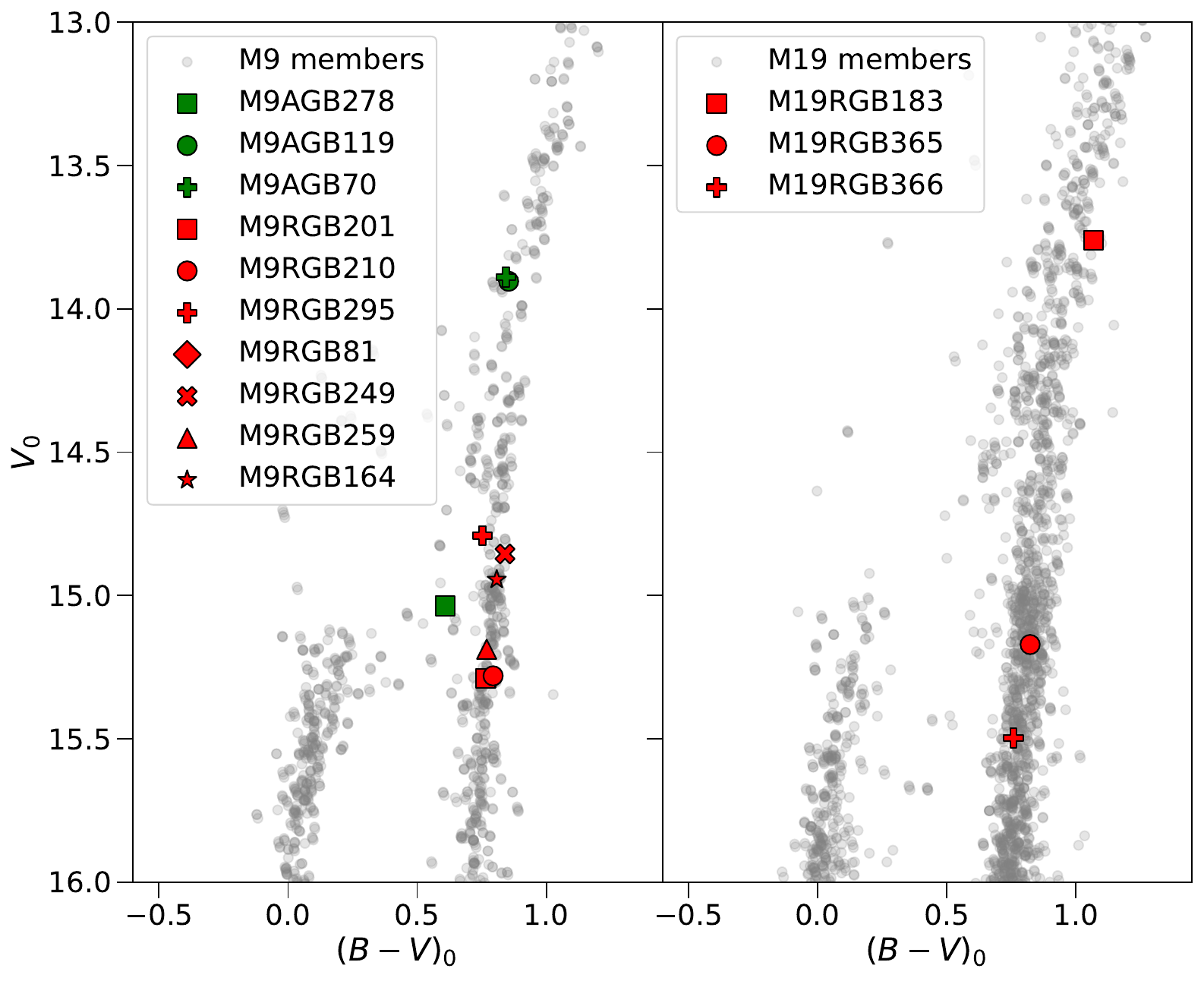}
	\caption{The reddening corrected colour-magnitude diagrams for M9 (left) and M19 (right), focusing on the magnitude range for our seismic sample. We illustrate the locations of the mass outliers by individual symbols, where we distinguish stars classified as RGB (red) and EAGB (green). Refer to Fig.~\ref{fig:isochrones} to see the full reddening correct colour-magnitude diagrams.}
	\label{fig:mass_outliers_CMD}
\end{figure}

Another explanation for the mass outliers is that we have misclassified their evolutionary phase due to differential reddening. Initially in Section~\ref{sec:data_prep}, we distinguished our sample into the RGB and EAGB evolutionary phases before corrections to the dust extinction were applied. Figure~\ref{fig:mass_outliers_CMD} shows the locations in the reddening corrected colour-magnitude diagram of the mass outliers in comparison to the cluster member sample. Interestingly, each star appears to be classified into the correct evolutionary phase\footnote{We noticed that the star M9AGB278 could be classified to be in the HB evolutionary phase from its location in the colour-magnitude diagram. However, it has a mass estimate ($0.54\pm0.04$(rand)$\pm0.01(sys)~\msun$) that is smaller than expected for the HB, and as such we are hesitant to change its evolutionary phase classification.}. Although, there is some uncertainty in the assignments of the evolutionary phases where the RGB and EAGB merge. Two stars in M9 (M9AGB70 \& M9AGB119) are located at this point where the two branches converge in the colour-magnitude diagram, and as such could be classified in either evolutionary phase. Additionally, both of these EAGB stars have masses that are consistent with the RGB average mass in Figure~\ref{fig:mass_outliers}, and therefore are potential candidates for being misclassified. Except for these stars, Figure~\ref{fig:mass_outliers_CMD} validates our method of classifying our sample into evolutionary phases in Section~\ref{sec:data_prep}, but eliminates it as an explanation for the majority of our mass outliers.

Instead, we suspect that their outlying masses are possibly due to incorrect measurement of the input parameters in the seismic mass relation. As shown in Figure~\ref{fig:mass_outliers}, the outlying masses are only found at the faint and bright ends of our samples. It is difficult to measure $\numax$ at these two extremes; faint stars with larger $\numax$ values have lower signal-to-noise and lower amplitudes of the solar-like oscillation, and bright stars with smaller $\numax$ values have a smaller number of excited modes and the solar-like oscillations are hard to differentiate from the granulation noise due to the limited length of the \textit{K2} data. We assigned stars that were observed to have lower quality in the power excess with a marginal detection (MD) quality flag (see Sec.~\ref{sec:seismic_params}), which can be identified by the open circles in Figure~\ref{fig:mass_outliers}. Approximately 70\% (7/10 stars for M9 and 2/3 for M19) of our mass outlier sample had a MD quality flag, implying that these stars could have an inaccurate measurement for $\numax$. Interestingly, the two EAGB stars that could be evolutionary phase misclassifications do not have the marginal detection quality flag. This further supports that their mass estimates are correct, and that they are actually RGB stars. 

In summary, although it would be exciting to describe our mass outliers as the product of a mass transfer event or merger, it is more likely that this inconsistency is attributed to our method; either in the incorrect classification of the evolutionary phase or inaccuracies in the measurement of the global asteroseismic $\numax$ parameter. An in-depth study of binaries in the red giant sample of M9 and M19 could alter this conclusion. 


\section{Summary}
\label{sec:conclusion}
Asteroseismology is one of the best tools to measure accurate individual stellar masses. Coupled with the \textit{relatively} homogeneous sample of stars in GCs, it is an excellent testbed to study stellar evolution. In this investigation, we expanded the number of seismically studied globular clusters to four, with detections of solar-like oscillations in two new GCs: M9 and M19. Using photometry from the \textit{K2} mission, we measured seismic masses for 55 stars in M9 and 37 stars in M19 in the RGB and EAGB evolutionary phases. 

To analyse the photometric data, we developed two different pipelines; \texttt{TPFstitch} and \texttt{pyMON}. \texttt{TPFstitch} can be used to assemble the cluster's TPFs into one product, called a TPF patch. This makes it easier to construct aperture masks for individual stars to extract their light curves. Currently, \texttt{TPFstitch} is only designed for M9 and M19, but has potential to be used for other objects observed in \textit{K2} (or possibly \textit{TESS} and other future photometric missions). Following the previous seismic GC studies (\citetalias{Howell22_M4} and \citetalias{Howell_M80}), we only compute masses using the $\Delta\nu$-independent scaling relation (Eq.~\ref{eq:mass_relation3}). This is because $\Delta\nu$ is difficult to measure with the low signal-to-noise and frequency resolution of our data. Our \texttt{pyMON} asteroseismic pipeline only measures $\numax$. Because there is no dependency on the poorly measured $\Delta\nu$ parameter, we found that the estimated $\numax$ uncertainty was reduced by a factor of 2 compared to previous pipelines. This in turn decreased the seismic mass uncertainty.

Using Figures~\ref{fig:M9_KDE} (M9) and \ref{fig:M19_KDE} (M19), we measured average masses for the RGB and EAGB evolutionary phases by taking the peak value of the KDEs (see Table~\ref{tab:average_masses}). The integrated mass loss was calculated as the difference between the average masses, and were found to be: $\Delta M_{\text{RGB-EAGB}} = 0.16\pm0.02$(rand){\raisebox{0.5ex}{\tiny$\substack{+0.03 \\ -0.03}$}}(sys)$~\msun$ (M9) and $\Delta M_{\text{RGB-EAGB}} = 0.33\pm0.03$(rand){\raisebox{0.5ex}{\tiny$\substack{+0.09 \\ -0.07}$}}(sys)$~\msun$ (M19). 

A distance modulus estimate was needed to calculate the seismic masses. In Section~\ref{sec:teffs_dust_lum}, we calculated our own distance modulus for each cluster, and determined values of $(m-M)_0 = 14.75 \pm0.13$ for M9 and $(m-M)_0 = 14.93 \pm0.17$ for M19. In Figure~\ref{fig:distmod_sys}, we demonstrate the effect on the average masses and mass loss when varying the distance modulus by $\pm2\sigma$. We found that even though the distance modulus has a significant systematic effect on the average masses, this effect is small ($10^{-2}~\msun$) for the mass differences (i.e the integrated mass loss). Therefore, asteroseismology can be used as an accurate method to measure the integrated mass loss for red giants in GCs despite the uncertainties in the distance modulus.

It is well understood that there is a metallicity dependence on the stellar mass loss; low metallicity stars will lose less mass compared to more metal-enriched stars. In Section~\ref{sec:massloss_metallicity}, we confirm this concept by comparing our mass loss estimates to the seismic measurements from two other GCs; M4 and M80 (Fig.~\ref{fig:massloss_metallicity}). We found that there was a discrepancy between the stellar mass loss of Type I clusters (homogeneous Fe abundances) and the Type II cluster (varying Fe abundances) M19. Our results suggest that stars in Type II clusters appear to loose more mass compared to stars in Type I clusters at a fixed metallicity. Excluding M19, we derived a \textit{preliminary} relationship between mass loss and metallicity for Type I cluster stars: $\Delta M_{\text{RGB-EAGB}} = 0.24~\text{[Fe/H]}+0.55$. Our relation is the first mass loss-metallicity trend that uses direct model-independent mass measurements. However to improve this relation, we need to populate Figure~\ref{fig:massloss_metallicity} with additional seismic mass loss measurements of other GCs at different metallicities.

We tested whether there is a mass difference between the light-element sub-populations by searching for a bimodal signature in our mass distributions. For M9, we did not observe a bimodality for the RGB sample in Figure~\ref{fig:M9_KDE}, and suspect that the sub-population mass difference is too small to detect (Sec.~\ref{sec:multi_pops}). In contrast, we have a weak detection of a bimodal mass distribution for our EAGB sample. However, the EAGB sample is small (5 stars), and thus this is not a conclusive detection of a mass difference between sub-population. To confirm whether this tentative bimodal modal mass distribution is correlated to sub-populations, we require spectroscopic abundance measurements of light elements. 

It is more complicated to study the light-element sub-populations for Type II clusters. These clusters are typically classified into Fe populations, which in turn contain their own sub-populations. To simplify our analysis, we assume that any detection of a mass difference would be due to metal-poor and metal-intermediate Fe populations (see Sec.~\ref{sec:multipops_M19}). The mass distributions for M19 in Figure~\ref{fig:M19_KDE} did not display any signatures of a mass bimodality in either the RGB or EAGB sample. However, the RGB mass distribution is broad, and as such there could be a hidden mass variation. Similar to our analysis for M9, we decompose the RGB KDE into two Gaussians, and measured a mass difference between the Fe populations of $0.13\pm0.03~\msun$. This is significantly larger than the measured mass difference between the Fe populations from isochrones ($\sim10^{-2}~\msun$). We suspect that the larger mass difference could be attributed to the combination of different mass signatures from the Fe populations and the light-element sub-populations. To disentangle the effect of the two types of populations in the mass distributions, we again need spectroscopy for our samples.

In Section~\ref{sec:mass_outliers}, we report the detection of ten mass outliers in M9 and three in M19 (Fig.~\ref{fig:mass_outliers}). We suggest that the outlying masses are  likely stars that have either been misclassified in their evolutionary status, or have inaccurate mass measurement. In Figure~\ref{fig:mass_outliers_CMD}, we tested the former theory using dust-corrected colour-magnitude diagrams. We concluded that only two stars from the mass outlier sample could have been incorrectly classified into their evolutionary phase: M9AGB70 \& M9AGB119. These EAGB stars have masses consistent with the average RGB mass for M9, and thus are likely RGB star candidates. For the rest of the sample, we suspect that inaccurate measurements of the $\numax$ parameter (as indicated by their marginal detection flag) resulted in the outlying masses. To confirm this, we would need better seismic data to remeasure $\numax$ for these stars. 

Finally, we would like to advocate for more time series photometric missions to observe GCs. This would require a bigger space based telescope and long-baseline observations to ensure that there is high enough signal-to-noise for the faint GC targets. Also, an instrument with small pixel sizes is needed to resolve stars in the high-density fields (preferably less than the scale of \textit{Kepler}'s: $4$"/pixel). These requirements could potentially be met with the upcoming \textit{Nancy Grace Roman Telescope} (\citealt{Akeson19_Roman_telescope}; expected launch in 2027), which represents the best opportunity to replicate this investigation on other GCs (see \citealt{Molnar_2023} for a detailed discussion). Additionally, to effectively study the multiple population phenomena in GCs we need to capitalise on the synergy between asteroseismology and spectropscopy for these stars. Again, this requires new high-resolution spectroscopic data for both clusters.


\section*{Acknowledgements}
S.W.C. acknowledges federal funding from the Australian Research Council through a Future Fellowship (FT160100046) and Discovery Projects (DP190102431 \& DP210101299). This research was supported by use of the Nectar Research Cloud, a collaborative Australian research platform supported by the National Collaborative Research Infrastructure Strategy (NCRIS). D.S. is supported by the Australian Research Council (DP190100666). C.K. acknowledges the `SeismoLab' KKP-137523 \'Elvonal grant of the Hungarian Research, Development and Innovation Office (NKFIH), and also thanks the hospitality of Monash University where part of this research was carried out.

Parts of this research was supported by the Australian Research Council Centre of Excellence for All Sky Astrophysics in 3 Dimensions (ASTRO 3D), through project number CE170100013.

This publication makes use of data products from the Two Micron All Sky Survey, which is a joint project of the University of Massachusetts and the Infrared Processing and Analysis Center/California Institute of Technology, funded by the National Aeronautics and Space Administration and the National Science Foundation. This paper includes data collected by the Kepler mission. Funding for the Kepler mission is provided by the NASA Science Mission directorate.  This work has also made use of data from the European Space Agency (ESA) mission {\it Gaia} (\url{https://www.cosmos.esa.int/gaia}), processed by the {\it Gaia} Data Processing and Analysis Consortium (DPAC,\url{https://www.cosmos.esa.int/web/gaia/dpac/consortium}). Funding for the DPAC has been provided by national institutions, in particular the institutions participating in the {\it Gaia} Multilateral Agreement.

\section*{Data Availability}

The main data in this article is available in Tables~\ref{tab:final_results_M9} and \ref{tab:final_results_M19}. The light curves will be shared on reasonable request to the author. 



\bibliographystyle{mnras}
\bibliography{refs} 




\appendix

\section{Full data table for M9}
\label{sec:Appendix_A}

\begin{figure}
	\centering
	\includegraphics[width=1\columnwidth]{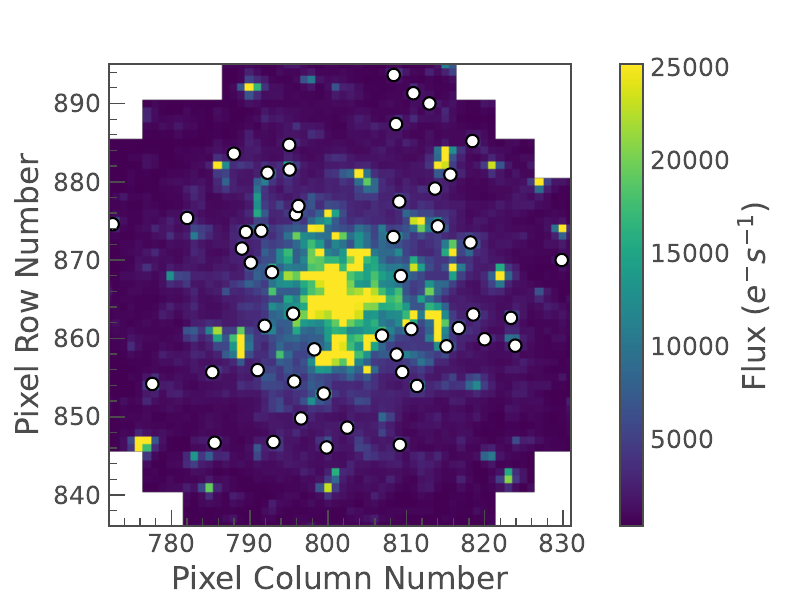}
	\caption{The M9 Superstamp created using \texttt{TPFstitch} (see Sec.~\ref{sec:superstamps}). White points indicate the locations of our seismic sample.}
	\label{fig:M9_Superstamp}
\end{figure}

Here we provide a table with our final results for the global seismic quantities, stellar properties, and mass estimates from Eq.~\ref{eq:mass_relation3} for our entire M9 sample. We also provide the cluster Superstamp in Figure~\ref{fig:M9_Superstamp} (see Sec.~\ref{sec:superstamps}).

\begin{table*}
\centering
\footnotesize
\caption{\small{We provide our own identifications (ID) and the corresponding \textit{Gaia} DR3 identifications for our M9 seismic sample. In the Campaign column we specify whether the full campaign or an individual sub-campaign light curve was used for that star. We have also included a quality flag (QF) for each star, where `MD' is a marginal detection and `D' is a true detection. Uncertainties for the luminosity and seismic mass are separated into random and systematic errors, respectively. The random uncertainties are included for $\log{(g)}$, and there was a constant systematic uncertainty of $\pm0.004$. We adopted a photometric $T_{\text{eff}}$ error is $\pm110\mathrm{K}$ for all stars (see Sec.~\ref{sec:teffs_dust_lum}). Stars with a bold ID have been identified as mass outliers (see Sec.~\ref{sec:mass_outliers}).} }
\begin{center}
\begin{tabular}{cccccccccc}
\hline
ID        & Gaia DR3 ID     & Campaign   & $\nu_{\text{max}}$ ($\mathrm{\mu Hz}$) &  QF & $T_{\text{eff}}$ & $L/L_{\odot}$       & $\log{(g)}$   & $M_{3}/M_{\odot}$      \\
\hline
M9AGB106 & 4134480384081650176 & Full & 3.9$\pm$0.2  & D  & 4710 & 266$\pm$10$\pm$5 & 1.58$\pm$0.02 & 0.69$\pm$0.07$\pm$0.01 \\
\textbf{M9AGB119} & 4134480384126602752 & Full & 4.8$\pm$0.1  & D  & 4729 & 243$\pm$9$\pm$5  & 1.67$\pm$0.01 & 0.76$\pm$0.07$\pm$0.02 \\
M9AGB126 & 4134480452836860672 & Full & 5.4$\pm$0.2  & MD & 4859 & 184$\pm$6$\pm$3  & 1.72$\pm$0.01 & 0.59$\pm$0.05$\pm$0.01 \\
M9AGB196 & 4122447294486921216 & Full & 12.3$\pm$0.3 & D  & 5354 & 113$\pm$2$\pm$2  & 2.05$\pm$0.01 & 0.59$\pm$0.04$\pm$0.01 \\
M9AGB257 & 4134480654674525312 & C11b & 10.8$\pm$0.5 & D  & 5111 & 123$\pm$3$\pm$2  & 2.01$\pm$0.02 & 0.66$\pm$0.06$\pm$0.01 \\
\textbf{M9AGB278} & 4122470693493013376 & Full & 18.1$\pm$0.4 & MD & 5394 & 73$\pm$1$\pm$1   & 2.22$\pm$0.01 & 0.54$\pm$0.04$\pm$0.01 \\
\textbf{M9AGB70}  & 4134456912095247488 & Full & 4.4$\pm$0.2  & D  & 4656 & 254$\pm$11$\pm$5 & 1.64$\pm$0.02 & 0.77$\pm$0.08$\pm$0.02 \\
M9AGB76  & 4134456946454967040 & Full & 8.1$\pm$0.2  & D  & 5064 & 166$\pm$4$\pm$3  & 1.88$\pm$0.01 & 0.68$\pm$0.05$\pm$0.01 \\
M9RGB103 & 4134480384117288448 & Full & 2.5$\pm$0.1  & MD & 4618 & 393$\pm$18$\pm$7 & 1.40$\pm$0.01 & 0.71$\pm$0.07$\pm$0.01 \\
M9RGB122 & 4134480414156301440 & Full & 5.6$\pm$0.1  & D  & 4784 & 202$\pm$7$\pm$4  & 1.73$\pm$0.01 & 0.70$\pm$0.06$\pm$0.01 \\
M9RGB130 & 4134480452836848640 & Full & 4.2$\pm$0.1  & D  & 4635 & 326$\pm$14$\pm$6 & 1.61$\pm$0.01 & 0.94$\pm$0.09$\pm$0.02 \\
M9RGB136 & 4134480482875841280 & Full & 3.1$\pm$0.1  & D  & 4678 & 322$\pm$13$\pm$6 & 1.49$\pm$0.01 & 0.68$\pm$0.06$\pm$0.01 \\
M9RGB137 & 4134480482875840128 & Full & 5.9$\pm$0.7  & D  & 4601 & 217$\pm$10$\pm$4 & 1.77$\pm$0.05 & 0.91$\pm$0.13$\pm$0.02 \\
\textbf{M9RGB164} & 4122446465543955584 & C11b & 18.0$\pm$0.2 & MD & 4828 & 90$\pm$3$\pm$2   & 2.24$\pm$0.01 & 0.98$\pm$0.08$\pm$0.02 \\
M9RGB185 & 4134456950774483072 & Full & 12.0$\pm$0.3 & D  & 4870 & 113$\pm$3$\pm$2  & 2.06$\pm$0.01 & 0.79$\pm$0.07$\pm$0.02 \\
M9RGB188 & 4134456980814711936 & Full & 17.6$\pm$0.2 & MD & 4849 & 87$\pm$3$\pm$2   & 2.23$\pm$0.01 & 0.91$\pm$0.08$\pm$0.02 \\
M9RGB189 & 4134456950741764736 & Full & 29.6$\pm$1.1 & MD & 5110 & 56$\pm$1$\pm$1   & 2.44$\pm$0.02 & 0.83$\pm$0.07$\pm$0.02 \\
M9RGB192 & 4122447255820159104 & Full & 11.1$\pm$0.2 & D  & 5033 & 135$\pm$3$\pm$3  & 2.02$\pm$0.01 & 0.78$\pm$0.06$\pm$0.02 \\
M9RGB195 & 4122447290179900928 & Full & 15.2$\pm$0.4 & D  & 4979 & 88$\pm$2$\pm$2   & 2.16$\pm$0.01 & 0.73$\pm$0.06$\pm$0.02 \\
M9RGB197 & 4122447290179903744 & Full & 9.7$\pm$0.3  & D  & 4826 & 130$\pm$4$\pm$2  & 1.97$\pm$0.01 & 0.76$\pm$0.07$\pm$0.02 \\
M9RGB198 & 4134456912095245056 & C11a & 18.8$\pm$0.6 & D  & 4865 & 66$\pm$2$\pm$1   & 2.26$\pm$0.01 & 0.73$\pm$0.06$\pm$0.02 \\
\textbf{M9RGB201} & 4134456912095242624 & C11a & 18.5$\pm$1.3 & MD & 4988 & 63$\pm$2$\pm$1   & 2.25$\pm$0.03 & 0.62$\pm$0.07$\pm$0.01 \\
M9RGB209 & 4122447290179904128 & Full & 17.2$\pm$0.2 & D  & 4844 & 80$\pm$3$\pm$2   & 2.22$\pm$0.01 & 0.82$\pm$0.07$\pm$0.02 \\
\textbf{M9RGB210} & 4134480384081644032 & Full & 17.4$\pm$0.5 & MD & 5146 & 61$\pm$2$\pm$1   & 2.21$\pm$0.01 & 0.51$\pm$0.04$\pm$0.01 \\
M9RGB214 & 4122470762199345408 & Full & 16.0$\pm$0.6 & MD & 4869 & 84$\pm$3$\pm$2   & 2.19$\pm$0.02 & 0.79$\pm$0.07$\pm$0.02 \\
M9RGB215 & 4134457156932926848 & C11b & 11.4$\pm$0.5 & D  & 4875 & 115$\pm$4$\pm$2  & 2.04$\pm$0.02 & 0.77$\pm$0.07$\pm$0.02 \\
M9RGB222 & 4134480654674447616 & C11b & 15.9$\pm$0.7 & MD & 4855 & 90$\pm$3$\pm$2   & 2.19$\pm$0.02 & 0.85$\pm$0.08$\pm$0.02 \\
M9RGB223 & 4134480654674445184 & Full & 19.8$\pm$0.4 & MD & 4818 & 77$\pm$3$\pm$1   & 2.28$\pm$0.01 & 0.93$\pm$0.08$\pm$0.02 \\
M9RGB224 & 4134480658995360768 & Full & 12.9$\pm$0.2 & D  & 4827 & 114$\pm$4$\pm$2  & 2.10$\pm$0.01 & 0.89$\pm$0.07$\pm$0.02 \\
M9RGB23  & 4122447294486893312 & Full & 7.2$\pm$0.5  & MD & 4754 & 167$\pm$6$\pm$3  & 1.84$\pm$0.03 & 0.76$\pm$0.09$\pm$0.02 \\
M9RGB24  & 4122447294486977920 & Full & 3.2$\pm$0.1  & D  & 4529 & 345$\pm$18$\pm$6 & 1.50$\pm$0.01 & 0.83$\pm$0.08$\pm$0.02 \\
M9RGB240 & 4134480418477086848 & C11b & 21.1$\pm$0.8 & MD & 4994 & 67$\pm$2$\pm$1   & 2.30$\pm$0.02 & 0.76$\pm$0.07$\pm$0.02 \\
M9RGB241 & 4134480418448226304 & Full & 19.8$\pm$0.3 & MD & 4856 & 76$\pm$2$\pm$1   & 2.28$\pm$0.01 & 0.89$\pm$0.07$\pm$0.02 \\
M9RGB242 & 4134480452836884736 & Full & 12.4$\pm$0.3 & D  & 4942 & 118$\pm$3$\pm$2  & 2.08$\pm$0.01 & 0.82$\pm$0.07$\pm$0.02 \\
M9RGB247 & 4134480654674526720 & Full & 14.8$\pm$0.3 & D  & 4970 & 105$\pm$3$\pm$2  & 2.15$\pm$0.01 & 0.85$\pm$0.07$\pm$0.02 \\
M9RGB248 & 4134480482875775104 & Full & 10.3$\pm$0.3 & D  & 4844 & 142$\pm$5$\pm$3  & 2.00$\pm$0.01 & 0.87$\pm$0.07$\pm$0.02 \\
\textbf{M9RGB249} & 4134480418477096320 & C11b & 14.8$\pm$0.5 & D  & 4690 & 103$\pm$4$\pm$2  & 2.16$\pm$0.02 & 1.02$\pm$0.10$\pm$0.02 \\
M9RGB250 & 4134480482875839744 & Full & 10.0$\pm$0.3 & MD & 5107 & 139$\pm$3$\pm$3  & 1.97$\pm$0.01 & 0.69$\pm$0.06$\pm$0.01 \\
M9RGB258 & 4134480654674525696 & Full & 8.7$\pm$0.7  & MD & 4781 & 143$\pm$5$\pm$3  & 1.93$\pm$0.03 & 0.78$\pm$0.09$\pm$0.02 \\
\textbf{M9RGB259} & 4134480689034264704 & Full & 24.2$\pm$0.8 & MD & 4840 & 72$\pm$2$\pm$1   & 2.37$\pm$0.01 & 1.04$\pm$0.09$\pm$0.02 \\
M9RGB264 & 4134480487167860480 & C11a & 21.1$\pm$0.5 & MD & 4921 & 78$\pm$2$\pm$1   & 2.31$\pm$0.01 & 0.93$\pm$0.08$\pm$0.02 \\
M9RGB281 & 4122470762212519936 & Full & 11.8$\pm$0.3 & D  & 4799 & 119$\pm$4$\pm$2  & 2.06$\pm$0.01 & 0.87$\pm$0.08$\pm$0.02 \\
M9RGB287 & 4134480418448220032 & C11b & 12.2$\pm$0.2 & D  & 4898 & 110$\pm$3$\pm$2  & 2.07$\pm$0.01 & 0.77$\pm$0.06$\pm$0.02 \\
M9RGB289 & 4134480521556296704 & C11b & 8.9$\pm$0.3  & D  & 4851 & 149$\pm$5$\pm$3  & 1.93$\pm$0.02 & 0.79$\pm$0.07$\pm$0.02 \\
M9RGB292 & 4122470899651473024 & Full & 14.7$\pm$0.3 & MD & 4797 & 95$\pm$3$\pm$2   & 2.15$\pm$0.01 & 0.86$\pm$0.07$\pm$0.02 \\
\textbf{M9RGB295} & 4122470659133217024 & Full & 20.9$\pm$0.3 & MD & 4976 & 99$\pm$3$\pm$2   & 2.30$\pm$0.01 & 1.13$\pm$0.09$\pm$0.02 \\
M9RGB35  & 4122470727852759552 & Full & 3.8$\pm$0.1  & MD & 4732 & 290$\pm$11$\pm$5 & 1.57$\pm$0.01 & 0.72$\pm$0.06$\pm$0.02 \\
M9RGB41  & 4122470895320181760 & Full & 7.2$\pm$0.2  & D  & 4799 & 162$\pm$6$\pm$3  & 1.84$\pm$0.01 & 0.72$\pm$0.06$\pm$0.02 \\
M9RGB54  & 4134456912095243392 & Full & 3.9$\pm$0.2  & MD & 4598 & 307$\pm$14$\pm$6 & 1.58$\pm$0.02 & 0.85$\pm$0.08$\pm$0.02 \\
M9RGB66  & 4134456916414734464 & C11b & 9.0$\pm$0.2  & MD & 4900 & 143$\pm$4$\pm$3  & 1.94$\pm$0.01 & 0.74$\pm$0.06$\pm$0.02 \\
M9RGB67  & 4134456912095237504 & C11b & 6.7$\pm$0.2  & D  & 4603 & 186$\pm$8$\pm$3  & 1.82$\pm$0.02 & 0.89$\pm$0.09$\pm$0.02 \\
M9RGB75  & 4134456946454970368 & Full & 7.5$\pm$0.3  & MD & 4830 & 159$\pm$5$\pm$3  & 1.86$\pm$0.02 & 0.72$\pm$0.06$\pm$0.02 \\
\textbf{M9RGB81}  & 4134456980814719616 & Full & 4.5$\pm$0.3  & MD & 4900 & 209$\pm$6$\pm$4  & 1.64$\pm$0.02 & 0.54$\pm$0.05$\pm$0.01 \\
M9RGB85  & 4134456980814708480 & Full & 4.7$\pm$0.1  & MD & 4581 & 246$\pm$12$\pm$5 & 1.67$\pm$0.01 & 0.84$\pm$0.08$\pm$0.02 \\
M9RGB88  & 4134456980814712320 & Full & 2.6$\pm$0.1  & D  & 4532 & 409$\pm$21$\pm$8 & 1.41$\pm$0.02 & 0.79$\pm$0.08$\pm$0.02\\
\hline
\label{tab:final_results_M9}
\end{tabular}
\end{center}
\end{table*}

\section{Full data table for M19}
\label{sec:Appendix_B}
This appendix provides a similar figure and table as Appendix~\ref{sec:Appendix_A}, but for M19.
\begin{figure}
	\centering
	\includegraphics[width=1\columnwidth]{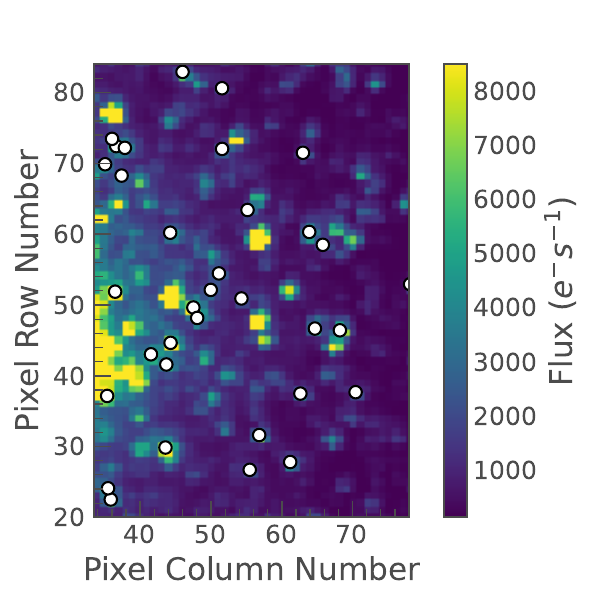}
	\caption{The M19 Superstamp created using \texttt{TPFstitch}. White points indicate the locations of our seismic sample.}
	\label{fig:M19_Superstamp}
\end{figure}

\begin{table*}
\centering
\footnotesize
\caption{\small{Same as Table~\ref{tab:final_results_M9} but for M19. Stars indicated by an asterisk have spectroscopic $\Teff$ measurements from APOGEE (see Sec.~\ref{sec:teffs_dust_lum}).} }
\begin{center}
\begin{tabular}{cccccccccc}
\hline
ID        & Gaia DR3 ID & Campaign        & $\nu_{\text{max}}$ ($\mathrm{\mu Hz}$) &  QF & $T_{\text{eff}}$ & $L/L_{\odot}$       & $\log{(g)}$   & $M_{3}/M_{\odot}$      \\
\hline
M19AGB17   & 4111914381107070720 & Full & 3.2$\pm$0.1  & D  & 4982 & 343$\pm$9$\pm$21  & 1.48$\pm$0.01 & 0.59$\pm$0.05$\pm$0.04 \\
M19AGB252  & 4111915171381128832 & Full & 4.9$\pm$0.3  & D  & 5019 & 179$\pm$5$\pm$11  & 1.66$\pm$0.02 & 0.46$\pm$0.04$\pm$0.03 \\
M19RGB275  & 4111914385460984192 & Full & 5.6$\pm$0.2  & D  & 5042 & 171$\pm$4$\pm$11  & 1.72$\pm$0.02 & 0.49$\pm$0.04$\pm$0.03 \\
M19AGB358  & 4111890879043608064 & C11a & 13.9$\pm$0.5 & D  & 5730 & 115$\pm$1$\pm$7   & 2.09$\pm$0.02 & 0.53$\pm$0.04$\pm$0.03 \\
M19RGB161* & 4111914316741433088 & Full & 3.4$\pm$0.1  & MD & 4732 & 359$\pm$14$\pm$22 & 1.52$\pm$0.01 & 0.78$\pm$0.07$\pm$0.05 \\
M19RGB162  & 4111914484186288384 & Full & 3.6$\pm$0.1  & D  & 4610 & 329$\pm$14$\pm$21 & 1.55$\pm$0.01 & 0.84$\pm$0.08$\pm$0.05 \\
\textbf{M19RGB183}* & 4111914454180426880 & Full & 3.5$\pm$0.1  & D  & 4497 & 362$\pm$20$\pm$23 & 1.54$\pm$0.01 & 0.97$\pm$0.10$\pm$0.06 \\
M19RGB198  & 4111915240100647936 & Full & 4.4$\pm$0.1  & D  & 4734 & 246$\pm$9$\pm$15  & 1.63$\pm$0.01 & 0.70$\pm$0.06$\pm$0.04 \\
M19RGB242* & 4111914415466818048 & Full & 5.7$\pm$0.1  & D  & 4769 & 230$\pm$8$\pm$14  & 1.74$\pm$0.01 & 0.83$\pm$0.07$\pm$0.05 \\
M19RGB247* & 4111891703679731840 & C11b & 7.2$\pm$0.2  & D  & 4897 & 194$\pm$6$\pm$12  & 1.84$\pm$0.01 & 0.80$\pm$0.07$\pm$0.05 \\
M19RGB258  & 4111914518546180736 & Full & 7.2$\pm$0.3  & D  & 4950 & 206$\pm$6$\pm$13  & 1.84$\pm$0.02 & 0.83$\pm$0.07$\pm$0.05 \\
M19RGB261* & 4111891669317603840 & Full & 8.0$\pm$0.3  & MD & 4876 & 179$\pm$6$\pm$11  & 1.89$\pm$0.02 & 0.84$\pm$0.07$\pm$0.05 \\
M19RGB271  & 4111914484186427136 & Full & 8.2$\pm$0.2  & MD & 4874 & 178$\pm$5$\pm$11  & 1.90$\pm$0.01 & 0.85$\pm$0.07$\pm$0.05 \\
M19RGB276* & 4111914282381730560 & Full & 6.8$\pm$0.3  & MD & 4891 & 184$\pm$6$\pm$11  & 1.81$\pm$0.02 & 0.72$\pm$0.06$\pm$0.05 \\
M19RGB288  & 4111914832114379008 & Full & 6.6$\pm$0.2  & D  & 4904 & 222$\pm$7$\pm$14  & 1.80$\pm$0.01 & 0.84$\pm$0.07$\pm$0.05 \\
M19RGB315  & 4111914312387596800 & Full & 9.2$\pm$0.1  & MD & 4844 & 161$\pm$5$\pm$10  & 1.95$\pm$0.01 & 0.89$\pm$0.07$\pm$0.06 \\
M19RGB331* & 4111914213638676096 & Full & 7.8$\pm$0.4  & MD & 4718 & 176$\pm$7$\pm$11  & 1.89$\pm$0.02 & 0.90$\pm$0.09$\pm$0.06 \\
M19RGB335  & 4111914110583003136 & Full & 8.7$\pm$0.2  & D  & 4829 & 161$\pm$5$\pm$10  & 1.93$\pm$0.01 & 0.85$\pm$0.07$\pm$0.05 \\
M19RGB336  & 4111915141375287808 & Full & 11.7$\pm$0.3 & D  & 4955 & 126$\pm$4$\pm$8   & 2.05$\pm$0.01 & 0.81$\pm$0.07$\pm$0.05 \\
M19RGB352* & 4111891570591766144 & Full & 11.9$\pm$0.3 & D  & 4940 & 128$\pm$4$\pm$8   & 2.06$\pm$0.01 & 0.85$\pm$0.07$\pm$0.05 \\
M19RGB355* & 4111891845469705216 & Full & 16.3$\pm$0.4 & MD & 4895 & 96$\pm$3$\pm$6    & 2.19$\pm$0.01 & 0.90$\pm$0.08$\pm$0.06 \\
M19RGB357  & 4111891639311243264 & C11a & 28.8$\pm$1.0 & MD & 5112 & 60$\pm$1$\pm$4    & 2.43$\pm$0.02 & 0.85$\pm$0.07$\pm$0.05 \\
M19RGB359  & 4111914346747325568 & Full & 14.5$\pm$0.3 & D  & 4994 & 104$\pm$3$\pm$6   & 2.14$\pm$0.01 & 0.81$\pm$0.06$\pm$0.05 \\
M19RGB360  & 4111915102688541184 & Full & 16.3$\pm$0.4 & D  & 5039 & 83$\pm$2$\pm$5    & 2.19$\pm$0.01 & 0.71$\pm$0.06$\pm$0.04 \\
\textbf{M19RGB365}  & 4111915175735656064 & Full & 14.1$\pm$0.9 & MD & 4924 & 83$\pm$2$\pm$5    & 2.13$\pm$0.03 & 0.66$\pm$0.07$\pm$0.04 \\
\textbf{M19RGB366}  & 4111915171381126656 & C11b & 22.7$\pm$0.6 & MD & 5095 & 59$\pm$1$\pm$4    & 2.33$\pm$0.01 & 0.67$\pm$0.05$\pm$0.04 \\
M19RGB367  & 4111915137021397248 & Full & 18.5$\pm$0.3 & D  & 5124 & 79$\pm$2$\pm$5    & 2.24$\pm$0.01 & 0.72$\pm$0.05$\pm$0.05 \\
M19RGB368  & 4111915205740870528 & Full & 17.1$\pm$0.6 & MD & 5155 & 87$\pm$2$\pm$5    & 2.20$\pm$0.01 & 0.72$\pm$0.06$\pm$0.05 \\
M19RGB389  & 4111914076223250176 & C11b & 25.9$\pm$1.3 & MD & 5058 & 65$\pm$2$\pm$4    & 2.39$\pm$0.02 & 0.87$\pm$0.08$\pm$0.06 \\
M19RGB392  & 4111914282381719936 & Full & 14.1$\pm$0.3 & D  & 4918 & 112$\pm$3$\pm$7   & 2.13$\pm$0.01 & 0.89$\pm$0.07$\pm$0.06 \\
M19RGB394  & 4111914381107070464 & Full & 21.1$\pm$0.5 & D  & 4975 & 76$\pm$2$\pm$5    & 2.30$\pm$0.01 & 0.88$\pm$0.07$\pm$0.06 \\
M19RGB395  & 4111914381107068800 & Full & 19.7$\pm$0.4 & D  & 4970 & 78$\pm$2$\pm$5    & 2.27$\pm$0.01 & 0.84$\pm$0.07$\pm$0.05 \\
M19RGB396  & 4111914381107069696 & Full & 12.7$\pm$0.8 & MD & 4934 & 102$\pm$3$\pm$6   & 2.08$\pm$0.03 & 0.72$\pm$0.07$\pm$0.05 \\
M19RGB399  & 4111914106229171840 & Full & 18.0$\pm$0.5 & D  & 4949 & 80$\pm$2$\pm$5    & 2.24$\pm$0.01 & 0.80$\pm$0.07$\pm$0.05 \\
M19RGB401  & 4111914209308389760 & Full & 24.8$\pm$0.8 & MD & 5035 & 70$\pm$2$\pm$4    & 2.37$\pm$0.02 & 0.91$\pm$0.08$\pm$0.06 \\
M19RGB403  & 4111914484186432000 & Full & 18.4$\pm$0.3 & D  & 5011 & 86$\pm$2$\pm$5    & 2.24$\pm$0.01 & 0.84$\pm$0.07$\pm$0.05 \\
M19RGB422  & 4111914827783854080 & C11b & 13.1$\pm$0.6 & MD & 5113 & 126$\pm$3$\pm$8   & 2.09$\pm$0.02 & 0.82$\pm$0.07$\pm$0.05\\
\hline
\label{tab:final_results_M19}
\end{tabular}
\end{center}
\end{table*}


\bsp	
\label{lastpage}
\end{document}